\documentclass[preprint,1pt,times]{elsarticle}

\usepackage{amssymb,amsmath,amsthm, amsfonts}
\usepackage{physics}
\usepackage{bbold}
\usepackage{bm}
\usepackage{rotating}
\usepackage{hyperref}
\usepackage{xcolor, color, soul}
\usepackage{caption}
\usepackage{subcaption}
\usepackage{enumitem}

\biboptions{sort&compress}

\journal{Physica A}

\begin{document}

\begin{frontmatter}

\title{Structural Controllability to Unveil Hidden Regulation Mechanisms in Unfolded Protein Response: the Role of Network Models}

\author[inst1,inst2]{Nicole Luchetti}

\affiliation[inst1]{organization={Engineering Department, Campus Bio-Medico University},
            addressline={Via Álvaro del Portillo, 21}, 
            city={Rome},
            postcode={00128}, 
            country={Italy}}

\author[inst1,inst2]{Alessandro Loppini\corref{cor}}
\ead{a.loppini@unicampus.it}
\author[inst1,inst3,inst4]{Margherita A. G. Matarrese}
\author[inst1]{Letizia Chiodo\corref{cor}}
\ead{l.chiodo@unicampus.it}
\author[inst1,inst5,inst6]{Simonetta Filippi}

\cortext[cor]{Corresponding authors}
\affiliation[inst2]{organization={Center for Life Nano- \& Neuro-Science, Istituto Italiano di Tecnologia},
            addressline={Viale Regina Elena, 291}, 
            city={Rome},
            postcode={00161}, 
            country={Italy}}

\affiliation[inst3]{organization={Jane and John Justin Institute for Mind Health Neurosciences Center, Cook Children Health Care System},
            addressline={1500 Cooper St.}, 
            city={Fort Worth},
            postcode={76104}, 
            state={Texas},
            country={USA}}
            
\affiliation[inst4]{organization={Department of Bioengineering, The University of Texas at Arlington},
            addressline={701 Nedderman Drive St.}, 
            city={Arlington},
            postcode={76019}, 
            state={Texas},
            country={USA}}
        
\affiliation[inst5]{organization={National Institute of Optics, National Research Council},
            addressline={Largo Enrico Fermi, 6}, 
            city={Florence},
            postcode={50125}, 
            country={Italy}}
            
\affiliation[inst6]{organization={International Center for Relativistic Astrophysics Network},
            addressline={Piazza della Repubblica, 10}, 
            city={Pescara},
            postcode={65122}, 
            country={Italy}}

\begin{abstract}

The Unfolded Protein Response is the cell mechanism for maintaining the balance of properly folded proteins in the endoplasmic reticulum , the specialized cellular compartment. Although it is largely studied from a biological point of view, much of the literature lacks a quantitative analysis of such a central signaling pathway. In this work, we aim to fill this gap by applying structural controllability analysis of complex networks to several Unfolded Protein Response networks to identify crucial nodes in the signaling flow. In particular, we first build different network models of the Unfolded Protein Response mechanism, relying on data contained in various protein-protein interaction databases. Then, we identify the driver nodes, essential for overall network control, i.e., the key proteins on which external stimulation may be optimally delivered to control network behavior. Our structural controllability analysis results show that the driver nodes commonly identified across databases match with known endoplasmic reticulum stress sensors. This potentially confirms that the theoretically identified drivers correspond to the biological key proteins associated with fundamental cellular activities and diseases. In conclusion, we prove that structural controllability is a reliable quantitative tool to investigate biological signaling pathways, and it can be potentially applied to networks more complex and less explored than Unfolded Protein Response.
\end{abstract}


\end{frontmatter}

\section{Introduction}\label{intro}

\noindent The Unfolded Protein Response (UPR) is a dynamic network of signaling processes in eukaryotic cells that adapts to various input signals. Activation of the UPR under endoplasmic reticulum (ER) stress conditions results in an adaptive response that prevents the increase of unfolded proteins via numerous pro-survival mechanisms, whose alterations may result in the formation of unfolded or misfolded proteins. This condition triggers the UPR, the adaptive process able to restore and maintain the cell's functions \cite{Hetz2012,SHEN2002,SHEN2005,Almanza,Walter2011,HETZ2018,GESSNER2014}. \newline

Binding immunoglobulin protein (BiP), also known as heat shock 70 kDa protein 5 (HSPA5), a molecular chaperone located in the ER lumen, is forced to dissociate from the luminal domain (LD) of the three ER stress transducers after the accumulation of unfolded or misfolded proteins. Therefore, HSPA5 triggers the dimerization of the ER stress sensors, and hence the activation of the cascading signaling pathways. As a result, the stress response begins with HSPA5 conveying the presence of unfolded proteins to the ER stress sensors; hence, HSPA5 can be regarded as the primary sensor that acts in the activation of UPR \cite{Ni2009, SOMMER20021, KOPP2019}. \newline

\begin{figure}[h!]
    \centering
    \addtolength{\leftskip}{-5cm}
    \addtolength{\rightskip}{-5cm}
    \includegraphics[width=1.35\textwidth]{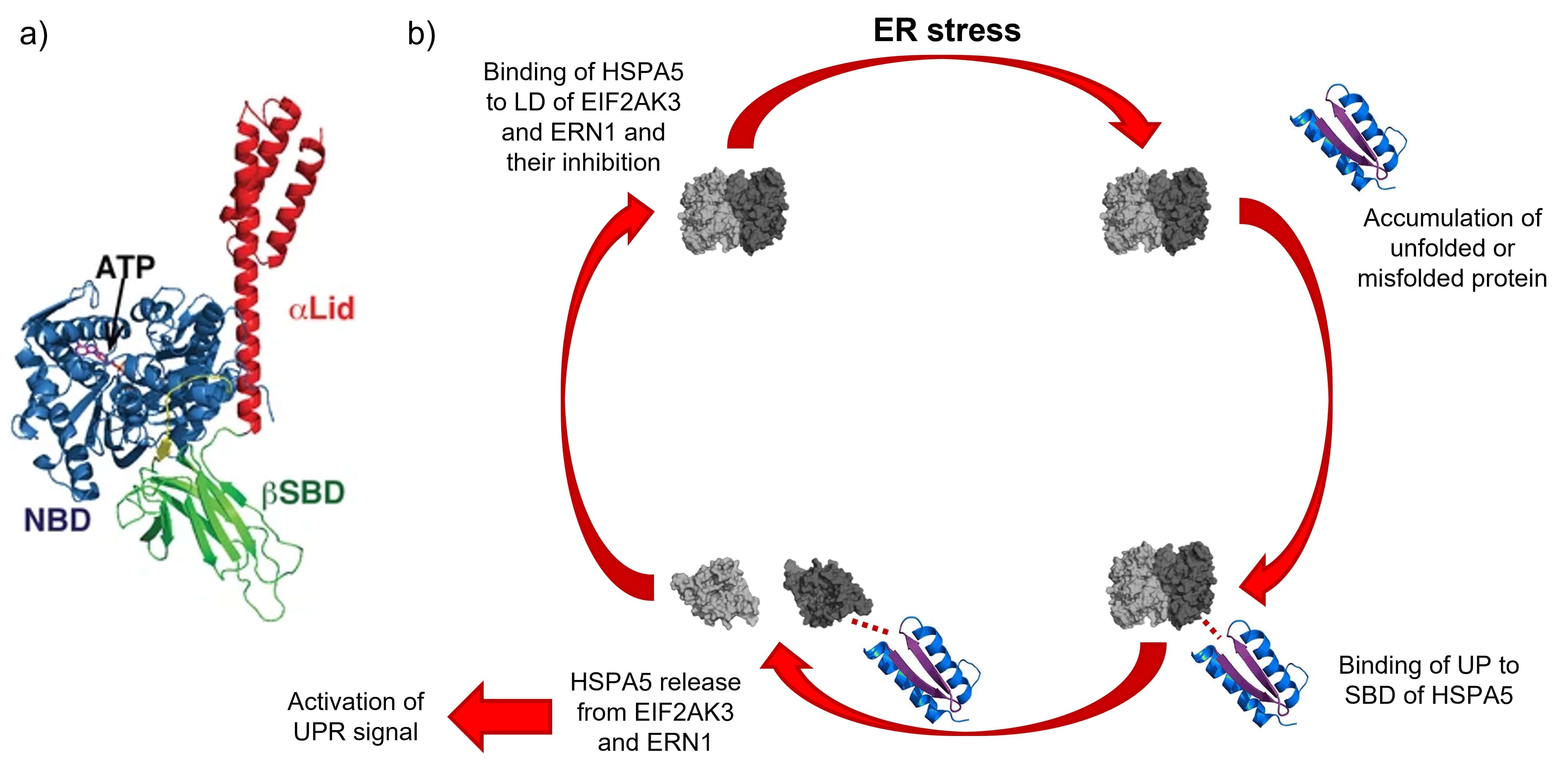}
    \caption{a) Molecular structure of HSPA5, from Ref. \cite{10.7554/eLife.29430} (NBD = nucleotide binding domain, SBD = substrate binding domain); b) Allosteric regulation of UPR activation, from Ref. \cite{Carrara2015} and \url{http://walterlab.ucsf.edu/upr-and-ire1/} (Reproduction under the Creative Commons Attribution License).}
    \label{UPR_activation}
\end{figure}

If the UPR adaptive response to ER stress fails to restore cell's homeostasis due to a persistent accumulation of unfolded proteins, the signaling program continues, and the ER stress sensors EIF2AK3 and ERN1 both drive multiple signaling outputs leading to apoptosis \cite{SHORE2011143, ANTHONY2012135, ijms20235896, HETZ2018169, FRIBLEY2009, Bartoszewska2020, KIM2006}; this signaling pathway is known as “terminal UPR”, and caspase enzymes play a central role in the transduction of ER apoptotic signals \cite{LI2008, McIlwain2013}.\newline

The activation and malfunctioning of UPR actors appears to be directly involved in protein misfolding diseases, according to experimental findings. Indeed, ER stress mediators have been considered in the description of a variety of neurodegenerative disorders, including Alzheimer's, Parkinson's, and Huntington's diseases, as well as prion and amyotrophic lateral sclerosis \cite{Scheper2015, Vidal2012, ijms21176127, biom10081090}. \newline

Since it is possible to describe the UPR mechanism as a network, in which the proteins that take part in the stress response and consequent apoptosis are identified by the nodes, and the protein-protein interactions are identified as the edges that connect proteins in pairs, in this work we try to apply the control theory of complex network proposed by Liu et al. \cite{Liu2011_1, RevModPhys.88.035006}, and based on Kalman’s criterion of controllability \cite{Kalman1963} for linear time-invariant systems (LTI). The state of a deterministic system, which is identified by the set of values of all the state variables (in the case of LTI systems, they are the state matrix and the control matrix or vector), completely describes the system at any time. 
The basic assumption behind the controllability is that a dynamical system may be steered from its initial state to any desired final state in a limited amount of time with the proper combination of input signals. In the last decades, the controllability of directed and weighted networks has been widely employed for a wide range of complex networks, including biological networks related to some interesting mechanisms involving protein-protein interactions (PPI) \cite{Uhart2016,Vinayagam2016,Abdallah2011,Ackerman2019,Kanhaiya2017,LIU2020122772}. Lin \cite{1100557} pioneered the concept of structural characteristics, which was later expanded by Shields and Pearson \cite{1101198} and alternatively derived by Glover and Silverman \cite{1101257}. 
In this framework, system parameters are either independent free variables or fixed zeros. This is consistent for physical systems’ models since parameter values related to nodes' interactions are never known exactly, except for zero values indicating the absence of interactions. In the case of PPI dynamical systems, experimental parameter values can be stored in particular databases, which can then be referenced to reconstruct the interaction matrix to be regulated. \newline

Here, we propose a controllability analysis of several Unfolded Protein Response
network models based on existing literature. We investigate multiple UPR models because it is at present impossible to uniquely identify a complete model for the mechanism. Using biological information from various protein-protein interaction databases, we test the network models and evaluated node ranking scores. Depending on the investigated database, the stored information may be vastly diverse and insufficient. Therefore, we demonstrate that the identification of driver nodes is, to some extent, dependent on both the network and the used database.

\section{Materials and Methods} \label{matmet}

\noindent Below we report the network models applied in this study, followed by the structural controllability and the minimum driver nodes methods, the network topological characterization, and lastly the statistical tools used to characterize the network analysis results.

\subsection{Network Models}\label{netmod}
Based on existing literature regarding UPR interactomes, we build different human Unfolded Protein Response networks, including both UPR \cite{Sengupta2019, Hetz2012, Huang2021, Erguler2013, Grootjans2016, Adams2019, Peng2021, Schroder2005, Halliday2014TargetingTU, Scheuner2008, Walter2011, Madden2019, Almanza, LEBEAUPIN2018927, HETZ2018169, YANG2020100860, READ2021, chen2019, clavarino2012, Chen2019bis,Lisbona2009,Carrara2017,Kim2008,Deldicque2013,Basseri2012,Junjappa2018,Li2016ATF6B} and apoptosis \cite{HOCKER2013,Sarosiek2013,CHONG2020537,Luo2013,Wei2020,Adams2019bis,garrison2012,ren2010, Gupta2016, Ghosh, HAN200716223, Cazanave2010, Erlacher2006, Ruckert2010, song2009, Siegmund2022, liu2011, MADKOUR2021113216, HATAI200026576, Xu2015, liu2004, wang2021, Shibue2003, Mesquida2019, Haneda2004, Kim2008} pathways.
We use text mining to reconstruct reliable networks with directional connections, because the protein-protein interactions involved in the in the URP and apoptosis pathways are not fully understood. We investigate the following networks: (i) 10 nodes (N10); (ii) 15 nodes (N15); (iii) network from Kalathur et al. \cite{Kalathur2012} (N26); and (iv) UPR + apoptosis network (N34). The N34 contains N15 which includes N10, while N26 shares 13 nodes with N34.

We select and compare the following 4 protein-protein interaction databases to build and analyze the network models \cite{Lehne2009, BAJPAI2020103380}, including direction and weight of each connection: (i) GPS-Prot (GPS; \url{http://gpsprot.org}) \cite{gpsprot}; (ii) Mentha (Men; \url{https://mentha.uniroma2.it/index.php}) \cite{mentha}; (iii) Signor 3.0 (Sig; \url{https://signor.uniroma2.it/}) \cite{signor}; and (iv) String (Str; \url{https://string-db.org/}) \cite{string1, string2}. We use the BioGRID database (\url{https://thebiogrid.org/}) \cite{biogrid} to add information on the direction of interactions which is occasionally unavailable from some of the selected databases. Moreover, we build refined networks models by recovering directly information from literature. These models are referred here as reference models (abbreviated in mrf). Hence, here we investigate a total of 19 network models. Each model is uniquely identified with a name containing the number of nodes and the interaction source (e.g. N10\textsubscript{GPS}, N15\textsubscript{mrf}).

\begin{figure}[h!]
    \centering
    \addtolength{\leftskip}{-5cm}
    \addtolength{\rightskip}{-5cm}
    \includegraphics[width=0.9\textwidth]{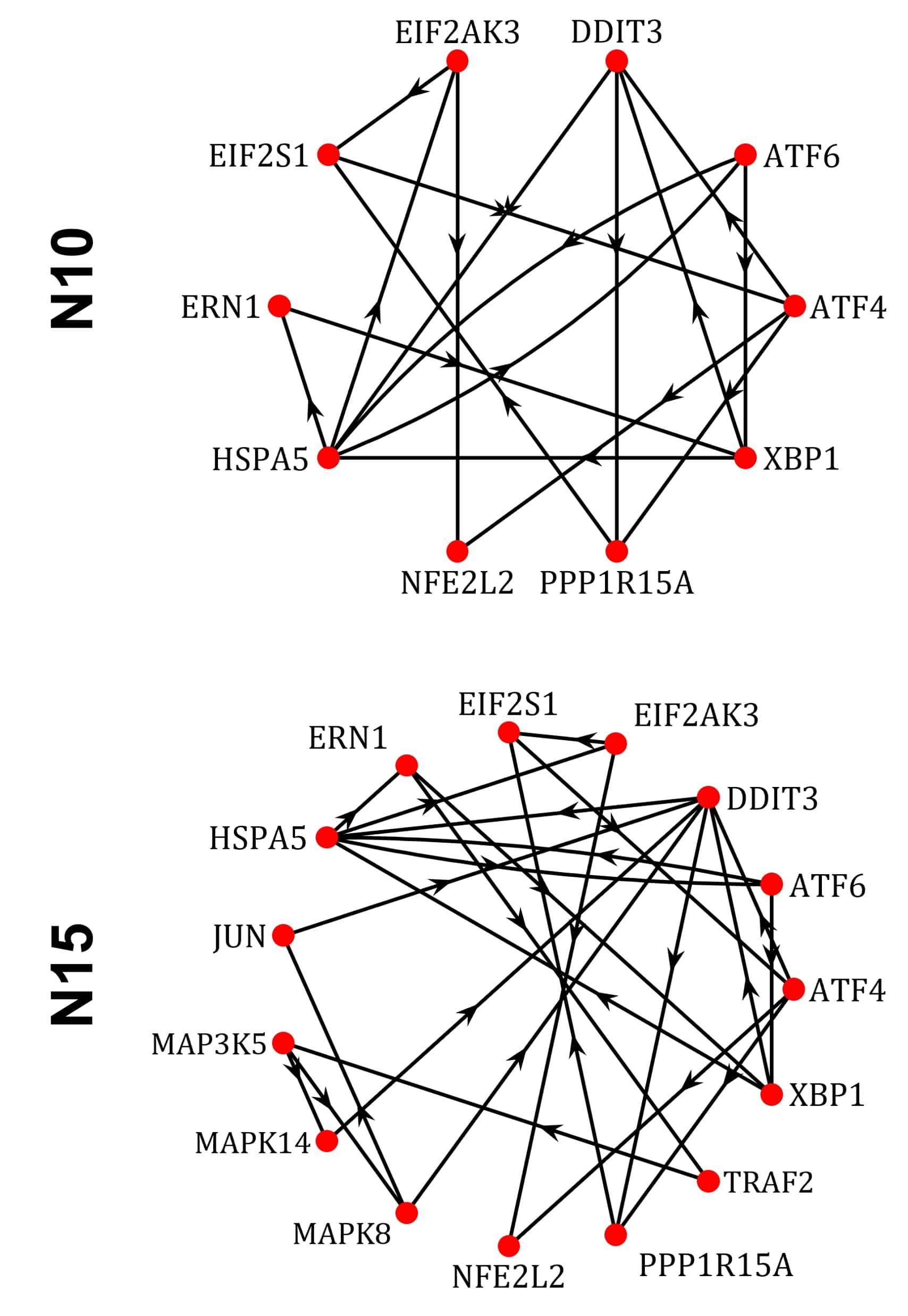}
    \caption{N10\textsubscript{mrf} and N15\textsubscript{mrf} UPR network models graph representation.}
    \label{UPR10_15_scheme}
\end{figure}

\paragraph{\textbf{UPR 10 Nodes Network - N10}}\label{UPR10}
The first explored network is the simplest one capable of describing the UPR, and it is formed by integrating the three primary pathways involved in the ER's stress response. The proteins included as network nodes are: eukaryotic translation initiation factor 2-alpha kinase 3 (EIF2AK3), activating transcription factor 6 (ATF6), inositol-requiring enzyme 1 $\alpha$ (ERN1), heat shock 70 kDa protein 5 (HSPA5), eukaryotic translation initiation factor 2 subunit 1 (EIF2S1), activating transcription factor 4 (ATF4), nuclear factor erythroid-derived 2-like 2 (NFE2L2), x-box binding protein 1 (XBP1), DNA damage-inducible transcript 3 (DDIT3), and protein phosphatase 1 regulatory subunit 15A (PPP1R15A). We report in Fig. \ref{UPR10_15_scheme} the N10\textsubscript{mrf} network model in graph representation, where the nodes are identified by red circles, and edges by directed black lines.

\paragraph{\textbf{UPR 15 Nodes Network - N15}}\label{UPR15}
N15 has five more proteins than N10 that are involved in the three main pathways of the stress response, namely: TNF receptor-associated factor 2 (TRAF2), mitogen-activated protein kinase 5 (MAP3K5), mitogen-activated protein kinase 8 (MAPK8), mitogen-activated protein kinase 14 (MAPK14), and transcription factor Jun (JUN). We report in Fig. \ref{UPR10_15_scheme} the N15\textsubscript{mrf} network model in graph representation, as in the previous case.

\paragraph{\textbf{Kalathur's Network - N26}}\label{UPR_Kal}
The N26 network is built based on the UPR model proposed by Kalathur and colleagues (2012). The full list of nodes includes (see Tab. 1 and Fig. 5 of Ref. \cite{Kalathur2012}): ATF4, ATF6, ATF6B, BAK1, BCL6, COPS5, CREB3L3, DDIT3, DNAJC3, EIF2AK3, EIF2S1, ERN1, GTF2I, GTPBP2, HSP90AA1, HSP90B1, HSPA5, MAP3K5, NFE2L2, NFYC, PPP1R15A, PSEN1, SRF, TAOK3, XBP1, and YY1. The interactome for EIF2AK3, ATF6, ERN1, and ATF4 is built using the Unified Human Interactome (UniHI) database (\url{www.unihi.org}) \cite{chaurasia}.

\paragraph{\textbf{UPR + Apoptosis Network Model - N34}}\label{Complete_UPR}
For a reliable complete model of UPR coupled to the apoptosis pathway, we build a network including information references \cite{Sengupta2019,Hetz2012,Huang2021,Erguler2013,Grootjans2016,Adams2019,Peng2021,Schroder2005,Halliday2014TargetingTU,Scheuner2008,Walter2011,Madden2019,Almanza,LEBEAUPIN2018927,HETZ2018169,YANG2020100860,READ2021,chen2019,clavarino2012,Chen2019bis,Lisbona2009,Carrara2017,Kim2008,Deldicque2013, Basseri2012, Junjappa2018, Li2016ATF6B, HOCKER2013, Sarosiek2013, CHONG2020537, Luo2013, Wei2020, Adams2019bis, garrison2012, ren2010, Gupta2016, Ghosh, HAN200716223, Cazanave2010, Erlacher2006, Ruckert2010, song2009, Siegmund2022, liu2011, MADKOUR2021113216, HATAI200026576, Xu2015, liu2004, wang2021, Shibue2003, Mesquida2019, Haneda2004, Kim2008}. Starting from N15, the N34 network has 19 more proteins, i.e. ATF6B, BAG5, BAK1, BAX, BBC3, BCL2, BCL2L11, BID, CASP1, CASP2, CASP8, IL1B, MAP2K7, MCL1, PMAIP1, TNFRSF10A, TNFRSF10B, TP53 and TXNIP.

\paragraph{\textbf{Ranking Scores Selection}}\label{Score} For each network, we perform all the analyses discarding the self-loop interactions included in GPS-Prot, Signor, and Mentha databases. We choose this approach because the inclusion of self-loop interactions could affected the nodes' ranking scores. However, String DB assigns different weights to different categories [i.e., text-mining, co-expression (CE), experimentally determined interactions (EXP), theoretical interactions (DB), homology (HOMOL), and so on]; thus, we reconstruct the total interaction weights using a weighted combination of CE, EXP, DB and HOMOL, with the following formula:
\begin{equation}
    \mathrm{String^{TOT}_{score} = 60\% EXP_{score} + 20\% CE_{score} + 10\% HOMOL_{score} + 10\% DB_{score}}.
\end{equation}

\subsection{Methodological Approach}\label{theomet}
\paragraph{\textbf{Structural Controllability}}
Using the four separate PPI databases for each of the four investigated networks, we employ the structural controllability theory to assess the nodes' ranking using the Kalman's rank condition for continuous linear time-invariant systems. Here, we provide a brief summary of the applied theory and redirect the reader to the original references for the complete theory. Generally, a linear differential equation can be used to describe the time-evolution of a network system consisting of N nodes, with M (M $\leq$ N) input signals acting on different nodes:

\begin{equation}
    \frac{d\textbf{x}(t)}{dt} = \textbf{A}\textbf{x}(t)+\textbf{B}\textbf{u}(t), 
\end{equation}

\noindent where $\textbf{x}$ = \{ $x_1$, $x_2$, ..., $x_N$\} is the state vector, $\textbf{u}$ = \{ $u_1$, $u_2$, ..., $u_M$\} is the input/control vector, \textbf{A} is a N$\times$N matrix representing the state matrix, and \textbf{B} is a N$\times$M matrix representing the input/control matrix. In \textbf{A}, each element identifies the interaction between two nodes, 

\begin{equation}
    \textbf{A} =
    \begin{pmatrix} 
        A_{11} & A_{12} & A_{13} & ... & A_{1N}  \\
        A_{21} & A_{22} & A_{23} & ... & A_{2N}  \\
        A_{31} & A_{32} & A_{33} & ... & A_{3N}  \\
        ... & ... & ... & ... & ...  \\
        A_{N1} & A_{N2} & A_{N3} & ... & A_{NN}
    \end{pmatrix}
\end{equation}
$ $\newline

Elements of \textbf{A} can be either 0 or 1, otherwise 0 or k $\in\mathbb{Q}^+$. In the first case, the matrix is called \textit{adjacency matrix}, and 1 identifies the presence of an interaction; in the second case, the matrix contains the biological information, extracted from databases, with a specific weight assigned for each interaction.\newline

In \textbf{B}, each element is the input signal acting on that node, and its form is

\begin{equation}
    \textbf{B} = 
    \begin{bmatrix}
        \textbf{e}_1 & \textbf{e}_2 & \textbf{e}_3 & ... & \textbf{e}_M
    \end{bmatrix}
\end{equation}

\noindent where \{$\textbf{e}_1, \textbf{e}_2, \textbf{e}_3, ..., \textbf{e}_{\mathrm{M}}$\} are the vectors of the canonical base. The dimensionality of \textbf{B} (i.e., M) depends on the number of input signals acting on the network. \newline

Given \textbf{A} and \textbf{B}, it is possible to assemble the controllability matrix, identified by \textbf{C}:

\begin{equation}
    \textbf{C} =
    \begin{bmatrix}
        \textbf{B} & \textbf{AB} & \textbf{A}^2\textbf{B} & \textbf{A}^3\textbf{B} & ... & \textbf{A}^{N-1}\textbf{B}
    \end{bmatrix}
\end{equation}

\noindent If the controllability matrix has a full rank, i.e., rank$(\textbf{C})=$ N, the network is fully controllable.

\paragraph{\textbf{Minimum Driver Nodes Identification}}
The Minimum Driver Nodes (MDN) algorithm proposed by Liu et al. \cite{Liu2011_1}, based on the minimal set of input signals required to fully control the network, and the MDN selection algorithm used by Zhang et al. \cite{8028698}, can be used to rank the nodes and then identify the driver nodes — the nodes on which an input signal must be injected to obtain full control of the network.
For each of the N nodes composing the network, the controllability matrix \textbf{C}$_i$ is evaluated as

\begin{equation}
    \textbf{C}_i =
    \begin{bmatrix}
        \textbf{B}_i & \textbf{AB}_i & \textbf{A}^2\textbf{B}_i & \textbf{A}^3\textbf{B}_i & ... & \textbf{A}^{N-1}\textbf{B}_i
    \end{bmatrix}
\end{equation}

\noindent where \textbf{B}$_\mathrm{i}$ = \textbf{e}$_\mathrm{i}$, i = 1, ..., N.
The nodes ranking scores are calculated as the rank of the controllability matrices \textbf{C}$_i$.
The minimum set of driver nodes (i.e., the number of input signals) is assessed by diagonalizing the state matrix, and by solving the formula

\begin{equation}
    n_{MDN} = \mathrm{max}_j[\beta(\lambda_j)] \equiv N - \mathrm{rank}(\lambda_{MAX} \mathbb{1}_N - \textbf{A})
\end{equation}

\noindent where $\lambda_j$ is the eigenvalues of \textbf{A}, $\lambda_{\mathrm{MAX}}$ is the eigenvalue  of \textbf{A} with the highest geometric multiplicity, which also identifies the minimum number of driver nodes, and $\mathbb{1}_\mathrm{N}$ is the N$\times$N identity matrix. \newline

For small networks, it is possible to precisely identify the set of MDN using an iterative process \cite{8028698}. Starting from the node with the highest ranking's score value and updating the input matrix \textbf{B}, the goal is to iteratively calculate the difference between the ranks of the controllability matrix including the input signal on an i$^{\mathrm{th}}$ node and the controllability matrix without the input signal:

\begin{equation}
    R_i = \mathrm{rank}(\lambda_{MAX} \mathbb{1}_N - \textbf{A}, [\textbf{B}, \textbf{e}_i]) - \mathrm{rank}(\lambda_{MAX} \mathbb{1}_N - \textbf{A}, \textbf{B})
\end{equation}

\noindent If R$_{\mathrm{i}}$ is bigger than 0, the node is a driver node. The calculation continues until the number of MDN is reached, and at each iteration \textbf{B}$_{\mathrm{new}}$ = [\textbf{B}$_{\mathrm{old}}$, \textbf{e}$_{\mathrm{new DN}}$].
In this specific application, the degree of the various networks, i.e. the number of involved nodes, is quite small; thus, the computational time is reasonable and, for example, the estimation time for the identification of DN for a set of about 10 nodes is few minutes. 

\paragraph{\textbf{Analysis of Topological Differences between Network Models}}\label{topolcon}
Recently, Ruan and colleagues \cite{Ruan2015} proposed a methodology for detecting interaction patterns in two-network comparisons. The algorithm relies on the Generalised Hamming Distance (GHD), which can be used for assigning the degree of topological difference between networks, and evaluating its statistical significance. Also, Mall et al. \cite{Mall2017} showed that the GHD statistic is more robust than other common topological measures, such as the Mean Absolute Difference (MAD) or the Quadratic Assignment Procedure (QAD). The GHD is thus a way to estimate the distance between two graphs and the lower the distance value, the lower the degree of topological difference between networks. \newline

If we consider two different networks, labeled U and V, with the same number of nodes (N), we can calculate the distance dGHD between the two networks as follows:

\begin{equation}
    dGHD(U,V) = \frac{1}{N(N-1)} \sum_{i \neq j} \big( u'_{ij} - v'_{ij} \big)^2 
\end{equation}

\noindent where $u'_{ij}$ and $v'_{ij}$ are mean centered edge-weights defined respectively as

\begin{equation}\begin{split}
    u'_{ij} = u_{ij} - \frac{1}{N(N-1)} \sum_{i \neq j} u_{ij}   \\
    v'_{ij} = v_{ij} - \frac{1}{N(N-1)} \sum_{i \neq j} v_{ij}
\end{split}\end{equation}

\noindent The edge weights $u_{ij}$ and $v_{ij}$ depend on the topology of the network and provide a measure of connectivity between every pair of i\textsuperscript{th} and j\textsuperscript{th} nodes in U and V, respectively. \newline

To assess whether the biological data stored in the reviewed PPI databases differ from each other, we apply the GHD theory to each of our network models.
Because the databases scores are assigned differently from database to database, it is only significant to compare the presence of pair interactions or not. As a quantitative test of information stored in databases, we evaluate the topological distance between our three reference models and each model obtained with the examined databases. In our case, lower values for the distance represent a greater consistency in the information between the literature and the databases.

\paragraph{\textbf{Statistical Analysis of Network Models}}\label{mlstat}
The implementation of structural controllability, the MDN finder and the GHD algorithm are performed using Pyhton 3.9 \cite{Rossum2009}. The statistical and graph analysis are performed using MATLAB R2022a (The MathWorks, Inc.) \cite{matlab2022}. The one-sample Kolmogorov-Smirnov test is used to determine the normality of the ranking score distributions for each UPR model. We apply non-parametric tests because our variables are not normally distributed. For paired comparisons between two given ranking scores across different databases considering one specific network, we use the Wilcoxon signed-rank test, and for non-paired comparisons between ranking score vectors across different networks considering one specific database, we use the Wilcoxon rank-sum test. If $p<0.01$, the results are regarded statistically significant. The Kruskal-Wallis test is used to evaluate multiple ranking score data vectors. If $p<0.05$, the results are regarded statistically significant. Finally, we report each ranking score distribution as a median and an IQR (between 25\% and 75\% percentile).

\section{Results} \label{resdisc}

\subsection{Structural controllability and topological analysis}
\paragraph{\textbf{N10 UPR Network Models}}\label{UPR10_resdisc}
We show the graphical representations of the N10 networks in Fig. \ref{1}, the nodes ranking scores and the number of MDN obtained for each model in Tab. \ref{tab1.1}. In Tab. \ref{tab1.2}, we report the MDN set for the five models of N10. Both weighted interactions and adjacency networks are used, to estimate the effect of biological information contained in the weighted interactions to properly describe the overall network behaviour. \newline

\begin{table}[ht!]
    \centering
    Adjacency matrix (weighted matrix)\\
    $ $\\
    \resizebox{0.75\textwidth }{!}{ 
    \begin{tabular}{c | ccccc }
    \hline
    \hline
    Node    & Reference & GPS-Prot & Mentha & Signor & String\\
    \hline
    ATF4    & 7 (-) & 5 (5)  &  2 (2) &  4 (5) &  6 (6)\\
    ATF6    & 9 (-) & 5 (5)  &  5 (5) &  6 (7) &  8 (9)\\
    DDIT3   & 8 (-) & 4 (4)  &  1 (1) &  2 (2) &  8 (8)\\
    EIF2AK3 & 9 (-) & 2 (2)  &  2 (2) &  6 (7) &  8 (8)\\
    EIF2S1  & 7 (-) & 1 (1)  &  1 (1) &  5 (6) &  6 (7)\\
    ERN1    & 9 (-) & 1 (1)  &  1 (1) &  1 (1) &  8 (9)\\
    HSPA5   & 8 (-) & 3 (3)  &  3 (3) &  6 (6) &  8 (8)\\
    NFE2L2  & 1 (-) & 1 (1)  &  1 (1) &  1 (1) &  1 (1)\\
    PPP1R15A& 8 (-) & 5 (5)  &  5 (5) &  1 (1) &  7 (8)\\
    XBP1    & 8 (-) & 4 (4)  &  4 (4) &  1 (1) &  7 (8)\\
    \hline
    \hline
    Eigenvalue & Reference & GPS-Prot & Mentha & Signor & String\\
    \hline
    0    & 2 (-) &  4 (4)  &  4 (4)  &  5 (4)  &  3  (2)\\
    \hline
    \end{tabular}}
    \caption{Nodes ranking scores and eigenvalues for N10 UPR models.}
    \label{tab1.1}
\end{table}

\begin{table}[h!]
    \centering
    \addtolength{\leftskip} {-5cm}
    \addtolength{\rightskip}{-5cm}
    \resizebox{1\textwidth }{!}{
    \begin{tabular}{c | c | c}
    \hline
    \hline
    Model & MDN set (adj.) & MDN set (weight.) \\
    \hline
    N10\textsubscript{mrf}    &   2: ATF6, EIF2AK3 & -\\
    N10\textsubscript{GPS}    &   4: ATF4, ATF6, EIF2AK3, PPP1R15A & 4: ATF4, ATF6, EIF2AK3, PPP1R15A\\
    N10\textsubscript{Men}    &   4: ATF4, ATF6, EIF2AK3, PPP1R15A  & 4: ATF4, ATF6, EIF2AK3, PPP1R15A\\
    N10\textsubscript{Sig}    &   5: ATF4, ATF6, EIF2AK3, NFE2L2, XBP1  & 4: ATF6, EIF2AK3, NFE2L2, XBP1 \\
    N10\textsubscript{Str}    &   3: ATF6, EIF2AK3, ERN1  & 2: ATF6, ERN1\\
    \hline
    \end{tabular}}
    \caption{MDN sets for N10 UPR models, referring to Tab. \ref{tab1.1}.}
    \label{tab1.2}
\end{table}

For GPS-Prot and Mentha DB models we find the same number of driver nodes for both adjacency and weighted networks (Tab. \ref{tab1.1}). This result may be interpreted as the number of active interactions of a high-ranked node being essential in predicting whether it can be a driving node or not. Two of the ER stress sensors, ATF6 and EIF2AK3, are identified as driver nodes in all databases. These findings are in agreement with the MDN set obtained using the reference model, for which we find two of the ER stress sensors as driver nodes. Still, we can expect that the number of driver nodes would be different because some of the information in N10\textsubscript{mrf} are not present in databases, while other interactions, absent in the literature, are found. For Signor DB, there is no stored information on possible NFE2L2 and XBP1 interactions during the UPR mechanism (Fig. \ref{1} - blue rows and columns - and the ranking scores equal to 1 in Tab. \ref{tab1.1}.

\begin{figure}[hb!]
   \centering
    \addtolength{\leftskip}{-5cm}
    \addtolength{\rightskip}{-5cm}
    \includegraphics[width=1.01\textwidth]{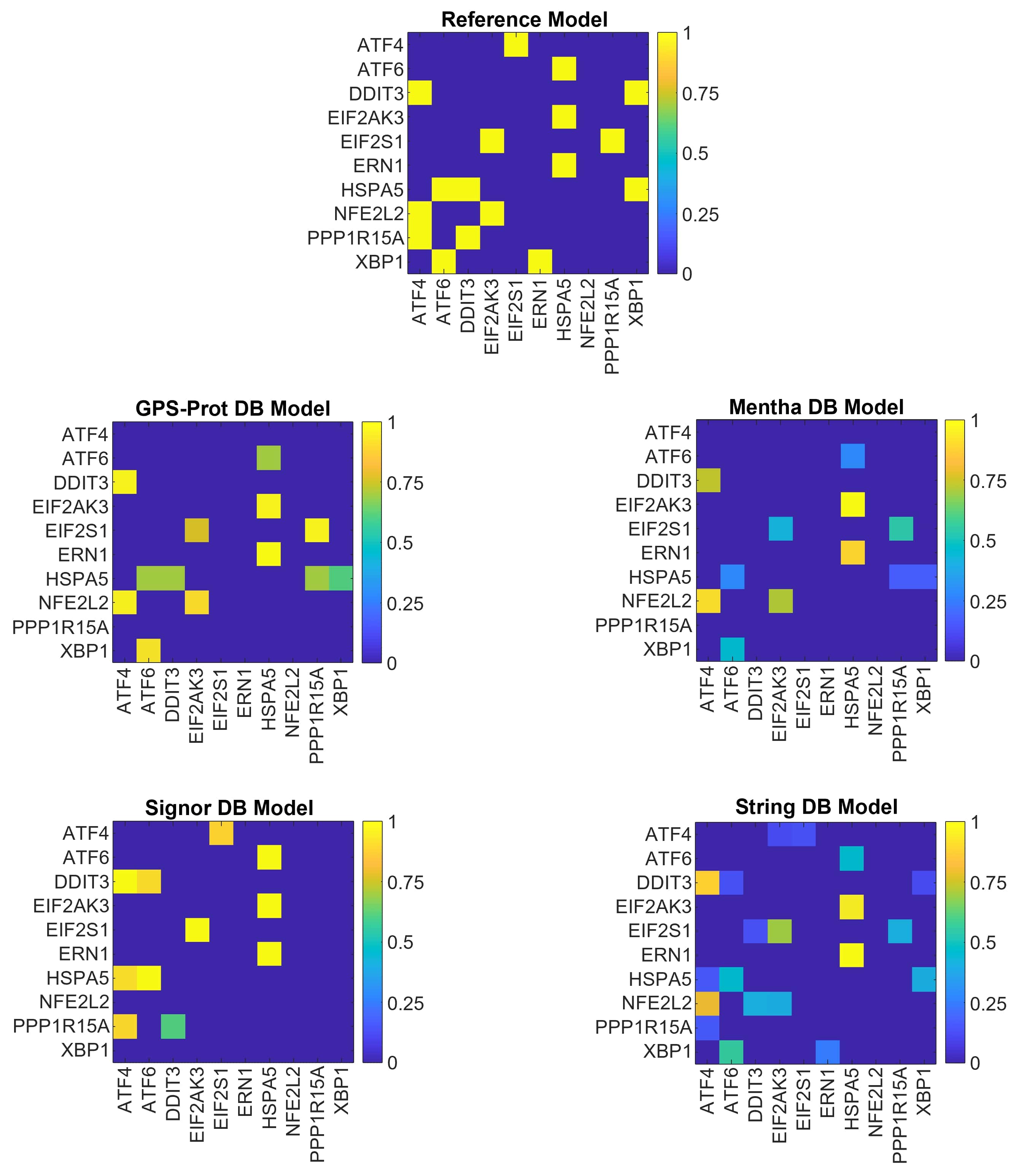}
    \caption{N10 UPR network models. The weights of interactions for each model are normalized with the maximum element of the matrix, and they are identified by the color bar. The (i, j) element of the matrix denotes the weight of the connection from j\textsuperscript{th} to i\textsuperscript{th} node.}
    \label{1}
\end{figure}

In Tab. \ref{tab1.3}, we report the dGHD for the topological comparison of all the possible networks pair. The MDN sets results are confirmed by the topological distance results, since the two closest models are the ones derived from the GPS-Prot and Mentha DBs. Thus, the information stored in these two databases is comparable and generated the same set of driver nodes. 
Finally, when we sort the distances of all the models from the reference (Tab. \ref{tab1.3} 2\textsuperscript{nd} column), we confirm that GPS-Prot stores more information than the others databases for defining the N10 (i.e. it had the minimum dGHD), whereas Signor is less informative (i.e. it shows the maximum dGHD).

\begin{table}[ht]
    \centering
    dGHD (adjacency matrix)\\
    $ $\\
    \begin{tabular}{c | ccccc}
    \hline
    \hline
    Model & N10\textsubscript{mrf} & N10\textsubscript{GPS} & N10\textsubscript{Men} & N10\textsubscript{Sig} & N10\textsubscript{Str}\\
    \hline
    N10\textsubscript{mrf}    &     -    & \textbf{0.065} & \textbf{0.075} & 0.107 & \textbf{0.077}\\
    N10\textsubscript{GPS}    &   \textbf{0.065}  &     -   & \textbf{0.011}  & 0.133 & 0.117\\
    N10\textsubscript{Men}    &   \textbf{0.075}  &   \textbf{0.011} & - & 0.122 & 0.104\\
    N10\textsubscript{Sig}    &   0.107  &   0.133 & 0.122  & - & 0.113\\
    N10\textsubscript{Str}    &   \textbf{0.077}  &   0.117 & 0.104  & 0.113 & -\\
    \hline
    \end{tabular}
    \caption{GHD distances between pairs of N10 models. The most similar models are  N10\textsubscript{Men}-N10\textsubscript{GPS}; the databases producing more similar results to the reference model are GPS-Prot, Mentha and String.}
    \label{tab1.3}
\end{table}

\paragraph{\textbf{N15 UPR Network Models}}\label{UPR15_resdisc}
As for the N10 network models, we show the graphical representations of the N15 networks in Fig. \ref{2}, the nodes ranking scores and the MDN sets obtained for each model in Tabs. \ref{tab2.1} and \ref{tab2.2}. In this case, we exclude String model from the analysis because the eigenvalue with the highest geometric multiplicity is not an integer value (see Tab. \ref{tab2.2}).\newline

The N15 models are different from the N10 ones mostly due to the additional nodes and thus the introduction of new interactions also for the nodes already present in N10. The nodes' ranking (Tab. \ref{tab2.1}) identifies more driver nodes candidates than in N10 and in this case GPS-Prot, Mentha and Signor give similar results for the adjacency and weighted matrices.
The N15 MDN sets (Tab. \ref{tab2.2}) also differ from the N10 ones (Tab. \ref{tab1.2}). All the possible explanations of obtained results are discussed in Sec. \ref{bio_disc}. \newline

In Tab. \ref{tab2.3}, we report the dGHD. The model closest to N15\textsubscript{mrf} is N15\textsubscript{Sig}. We find that the closest pair is N15\textsubscript{GPS} - N15\textsubscript{Men} that, as for N10 network, are the most distant models to the reference.

\begin{table}[h]
    \centering
    Adjacency matrix (weighted matrix)\\
    $ $\\
    \resizebox{0.75\textwidth }{!}{ 
    \begin{tabular}{c | ccccc}
    \hline
    \hline
    Node    & Reference & GPS-Prot & Mentha & Signor & String\\
    \hline
    ATF4  &  10 (-)  &   9 (9)   &  5 (5)  &  6 (6) &  10 (10)\\
    ATF6  &  13 (-)  &  11 (11)  & 10 (10) &  6 (6) &  13 (13)\\
    DDIT3 &  10 (-)  &   9 (5)   &  1 (1)  &  2 (2) &  10 (10)\\
    EIF2AK3& 12 (-)  &  10 (10)  &  6 (6)  &  8 (8) &  11 (12)\\
    EIF2S1&  10 (-)  &   1 (1)   &  1 (1)  &  7 (7) &  11 (11)\\
    ERN1  &  13 (-)  &  11 (11)  & 10 (10) &  4 (4) &  13 (13)\\
    HSPA5 &  11 (-)  &   8 (8)   &  8 (8)  &  6 (6) &  12 (12)\\
    JUN   &  10 (-)  &   8 (8)   &  4 (4)  &  3 (3) &  10 (10)\\
    MAP3K5&  11 (-)  &   1 (1)   &  1 (1)  &  1 (1) &  11 (11)\\
    MAPK14&  11 (-)  &  10 (10)  &  6 (6)  &  3 (3) &  10 (10)\\
    MAPK8 &  11 (-)  &   8 (8)   &  4 (4)  &  4 (4) &  10 (10)\\
    NFE2L2&   1 (-)  &   9 (9)   &  5 (5)  &  1 (1) &  11 (11)\\
    PPP1R15A& 11 (-) &  10 (10)  &  9 (9)  &  1 (1) &  12 (12)\\
    TRAF2 &   12 (-) &  10 (10)  &  9 (9)  &  2 (2) &  12 (12)\\
    XBP1  &   12 (-) &   9 (9)   &  9 (9)  &  1 (1) &  11 (11)\\
    \hline
    \hline
    Eigenvalue & Reference & GPS-Prot & Mentha & Signor & String\\
    \hline
    0    & 3 (-) &  4 (4)  &  4 (4)  &  6 (4)  &  -  (-)\\
    j$\in \mathbb{R}$, $\longrightarrow$ 0 & - (-)   &   - (-)  &  - (-)  &  - (-)  &  3 (3)\\
    \hline
    \end{tabular}}
    \caption{Nodes ranking scores and eigenvalues for N15 UPR models.}
    \label{tab2.1}
\end{table}

\begin{table}[h]
    \centering
    \addtolength{\leftskip} {-5cm}
    \addtolength{\rightskip}{-5cm}
    \resizebox{1.3\textwidth }{!}{
    \begin{tabular}{c | c | c}
    \hline
    \hline
    Model & MDN set (adj.) & MDN set (weight.) \\
    \hline
    N15\textsubscript{mrf}  &   3: ATF6, ERN1, MAPK14              & -\\
    N15\textsubscript{GPS}  &   4: ATF6, ERN1, MAPK14, PPP1R15A    &  4: ATF6, ERN1, MAPK14, PPP1R15A\\
    N15\textsubscript{Men}  &   4: ATF6, ERN1, MAPK14, PPP1R15A    &  4: ATF6, ERN1, MAPK14, PPP1R15A\\
    N15\textsubscript{Sig}  &   6: ATF6, EIF2AK3, MAPK14, MAPK8, NFE2L2, XBP1   &  6: ATF6, EIF2AK3, MAPK14, MAPK8, NFE2L2, XBP1\\
    \hline
    \end{tabular}}
    \caption{MDN sets for N15 UPR models, referring to Tab. \ref{tab2.1}.}
    \label{tab2.2}
\end{table}

\begin{table}[h!]
    \centering
    dGHD (adjacency matrix)\\
    $ $\\
    \begin{tabular}{c | ccccc}
    \hline
    \hline
    Model & N15\textsubscript{mrf} & N15\textsubscript{GPS} & N15\textsubscript{Men} & N15\textsubscript{Sig} & N15\textsubscript{Str}\\
    \hline
    N15\textsubscript{mrf}    &     -    &   0.099 & 0.104  & \textbf{0.065} & \textbf{0.084}\\
    N15\textsubscript{GPS}    &   0.099  &     -   & \textbf{0.005}  & 0.115 & 0.109\\
    N15\textsubscript{Men}    &   0.104  &   \textbf{0.005} &   -    & 0.111 & 0.104\\
    N15\textsubscript{Sig}    &   \textbf{0.065}  &   0.115 & 0.111  &   -   & 0.099\\
    N15\textsubscript{Str}    &   \textbf{0.084}  &   0.109 & 0.104  & 0.099 & -\\
    \hline
    \end{tabular}
    \caption{GHD distances between the N15\textsubscript{mrf} and the other four N15 UPR models. The most similar models are  N15\textsubscript{GPS}-N15\textsubscript{Men}; the databases producing more similar results to the reference model are Signor and String.}
    \label{tab2.3}
\end{table}

\clearpage
\begin{figure}[h!]
    \vspace*{-1.5 cm}   
    \centering
    \addtolength{\leftskip} {-5cm}
    \addtolength{\rightskip}{-5cm}
    \includegraphics[width=1.4\textwidth]{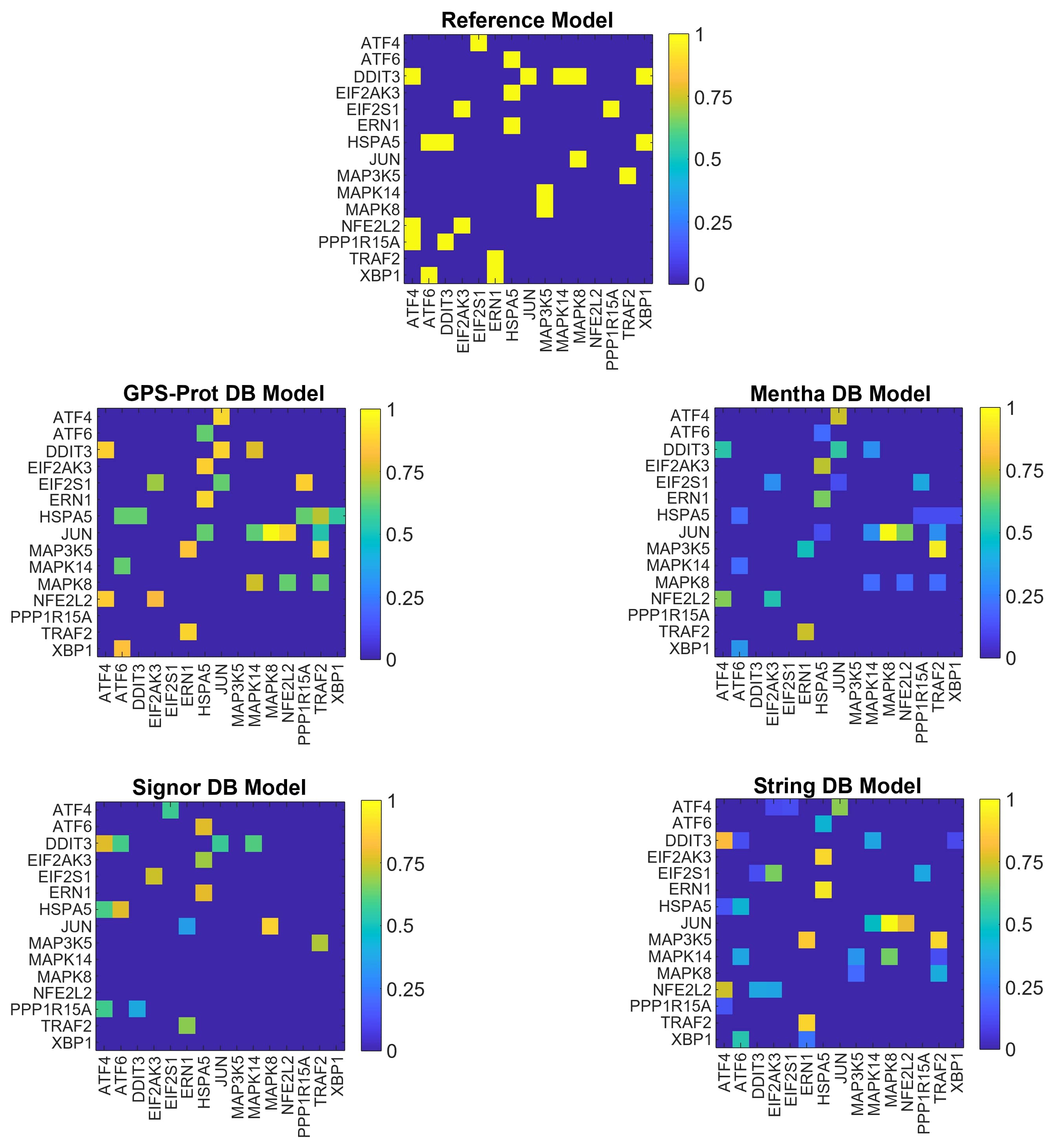}
    \caption{N15 UPR network models. The weights of interactions for each model are normalized with the maximum element of the matrix, and they are identified by the color bar. The (i, j) element of the matrix denotes the weight of the connection from j\textsuperscript{th} to i\textsuperscript{th} node.}
    \label{2}
\end{figure}

\begin{figure}[h!]
   \centering
    \addtolength{\leftskip} {-5cm}
    \addtolength{\rightskip}{-5cm}
   \includegraphics[width=1.5\textwidth]{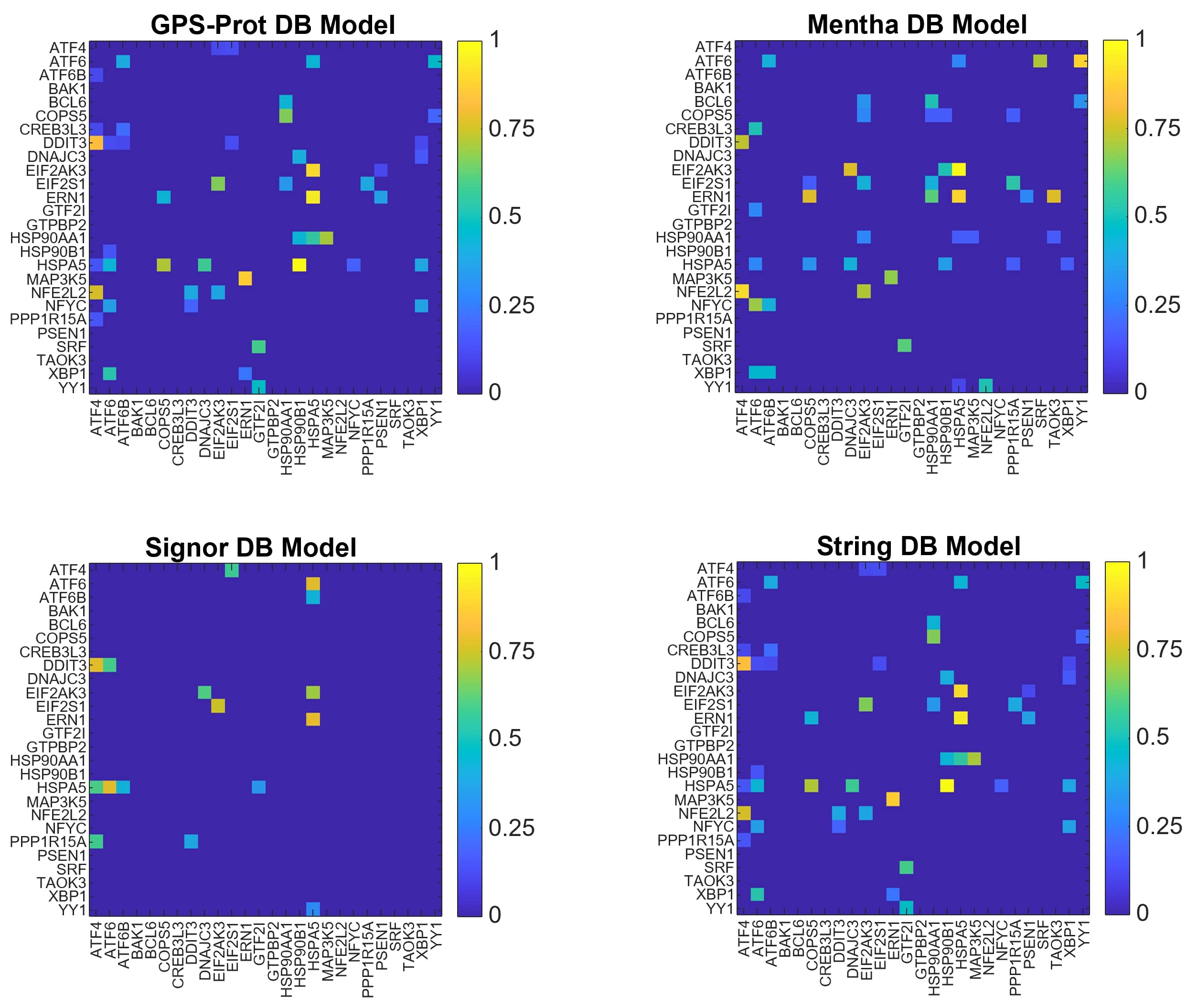}
    \caption{N26 UPR network models. The weights of interactions for each model are normalized with the maximum element of the matrix, and they are identified by the color bar. The (i, j) element of the matrix denotes the weight of the connection from j\textsuperscript{th} to i\textsuperscript{th} node.}
    \label{3}
\end{figure}

\paragraph{\textbf{N26 UPR Network Models}}\label{UPRKalathur_resdisc}
For N26, we show the graphical representations of the multiple network models in Fig. \ref{3}, the nodes ranking scores and the MDN sets obtained for each model in Tab. \ref{tab3.1}. Interestingly, the MDN sets for these networks are composed by an high number of proteins, compared to the previous cases (lower panel of Tab. \ref{tab3.1}). This finding may be explained considering that, for most of the nodes composing the N26 network, the biological information is lacking in the considered databases. With this respect, the representation of the adjacency network within Signor DB is quite explanatory, suggesting that Signor DB contains the lower quantity of information for this particular network.

\begin{table}[h!] 
\centering
    Adjacency matrix (weighted matrix)\\
    $ $\\
    \resizebox{0.7\textwidth }{!}{ 
    \begin{tabular}{c | cccc}
    \hline
    \hline
    Node & GPS-Prot & Mentha & Signor & String\\
    \hline
    ATF4    &   13 (13)  &  13 (13) &  5 (5) &  14 (14)\\
    ATF6    &   11 (11)  &  11 (11) &  6 (7) &  14 (14)\\
    ATF6B   &   13 (13)  &  14 (14) &  7 (7) &  15 (15)\\
    BAK1    &   12 (12)  &  1  (1)  &  1 (1) &   1 (1)\\
    BCL6    &    1 (1)   &  1  (1)  &  1 (1) &   1 (1)\\
    CPPS5   &   10 (10)  &  12 (12) &  1 (1) &  14 (14)\\
    CREB3L3 &    1 (1)   &  1  (1)  &  1 (1) &   1 (1)\\ 
    DDIT3   &   12 (12)  &  1  (1)  &  2 (2) &  14 (14)\\
    DNAJC3  &   13 (13)  &  13 (13) &  8 (8) &  13 (14)\\
    EIF2AK3 &   12 (12)  &  12 (12) &  7 (7) &  14 (14)\\
    EIF2S1  &   11 (11)  &  1  (1)  &  6 (6) &  14 (14)\\
    ERN1    &   11 (11)  &  12 (12) &  1 (1) &  13 (13)\\
    GTF2I   &   12 (12)  &  13 (13) &  7 (7) &  16 (16)\\
    GTPBP2  &   1  (1)   &  1  (1)  &  1 (1) &   1 (1)\\
    HSP90AA1&   10 (10)  &  10 (10) &  1 (1) &  14 (14)\\
    HSP90B1 &   13 (13)  &  13 (13) &  1 (1) &  15 (15)\\
    HSPA5   &   11 (11)  &  12 (12) &  6 (6) &  12 (13)\\
    MAP3K5  &   10 (10)  &  11 (11) &  1 (1) &  15 (15)\\
    NFE2L2  &    1 (1)   &  12 (12) &  1 (1) &   1 (1)\\
    NFYC    &    1 (1)   &  1  (1)  &  1 (1) &  13 (14)\\
    PPP1R15A&   12 (12)  &  13 (13) &  1 (1) &  15 (15)\\
    PSEN1   &   12 (12)  &  13 (13) &  1 (1) &  15 (15)\\
    SRF     &   12 (12)  &  12 (12) &  1 (1) &   1 (1)\\
    TAOK3   &   12 (12)  &  13 (13) &  1 (1) &   1 (1)\\
    XBP1    &   12 (12)  &  13 (13) &  1 (1) &  15 (15)\\
    YY1     &   12 (12)  &  11 (11) &  1 (1) &  15 (15)\\
    \hline
    \hline
    Eigenvalue & GPS-Prot & Mentha & Signor & String\\
    \hline
    0          & 11 (11)  &  11 (11) &  19 (19)  &  8 (8)\\
    \hline
    \end{tabular}}
\caption{Nodes ranking scores and eigenvalue for N26 UPR models.} 
\label{tab3.1} 
\end{table}

Another possible explanation for the poor results obtained for N26 is related to the database \cite{chaurasia} used by the authors to build the PPI network, that has not been updated since 2009 and is also, at present, inaccessible. Due to the high number of MDN, we do not identify the driver nodes because they may not be very significant. As it is not possible to reconstruct all the interactions as defined by the authors \cite{Kalathur2012}, we apply the theory for topological significance only on the models built on the four selected databases.

In Tab. \ref{tab3.2}, we report the pairwise dGHD for N26 newtork models. \newline
\noindent As expected, the smallest value of the distance is obtained with GPS-Prot and Mentha DBs. Indeed, the ranking scores obtained with N26\textsubscript{GPS} and N26\textsubscript{Men} models (first two columns of Tab. \ref{tab3.1}) are quite similar, and the number of MDN (lower panel of Tab. \ref{tab3.1}) was the same. It is interesting to note the fact that all the other evaluated distances have more comparable values than what is obtained for the other models. This may be because the ratio of stored information is similar for each pair of DBs regarding the network built by Kalathur and colleagues.

\begin{table}[h!]
    \centering
    dGHD (adjacency matrix)\\
    $ $\\
    \resizebox{0.6\textwidth }{!}{
    \begin{tabular}{c | cccc}
    \hline
    \hline
    Model & N26\textsubscript{GPS} & N26\textsubscript{Men} & N26\textsubscript{Sig} & N26\textsubscript{Str}\\
    \hline
    N15\textsubscript{GPS}    &   -      &   \textbf{0.009} & 0.072  & 0.062\\
    N15\textsubscript{Men}    &   \textbf{0.009}  &     -   & 0.069  & 0.066\\
    N15\textsubscript{Sig}    &   0.072  &   0.069 &   -    & 0.067\\
    N15\textsubscript{Str}    &   0.062  &   0.066 & 0.067  &   -  \\
    \hline
    \end{tabular}
    }
    \caption{GHD distances between the four N26 UPR models. The most similar models are  N26\textsubscript{GPS}-N26\textsubscript{Men}.}
    \label{tab3.2}
\end{table}

\begin{figure}[h!]
   \centering
    \addtolength{\leftskip} {-5cm}
    \addtolength{\rightskip}{-5cm}
    \includegraphics[width=1.3\textwidth]{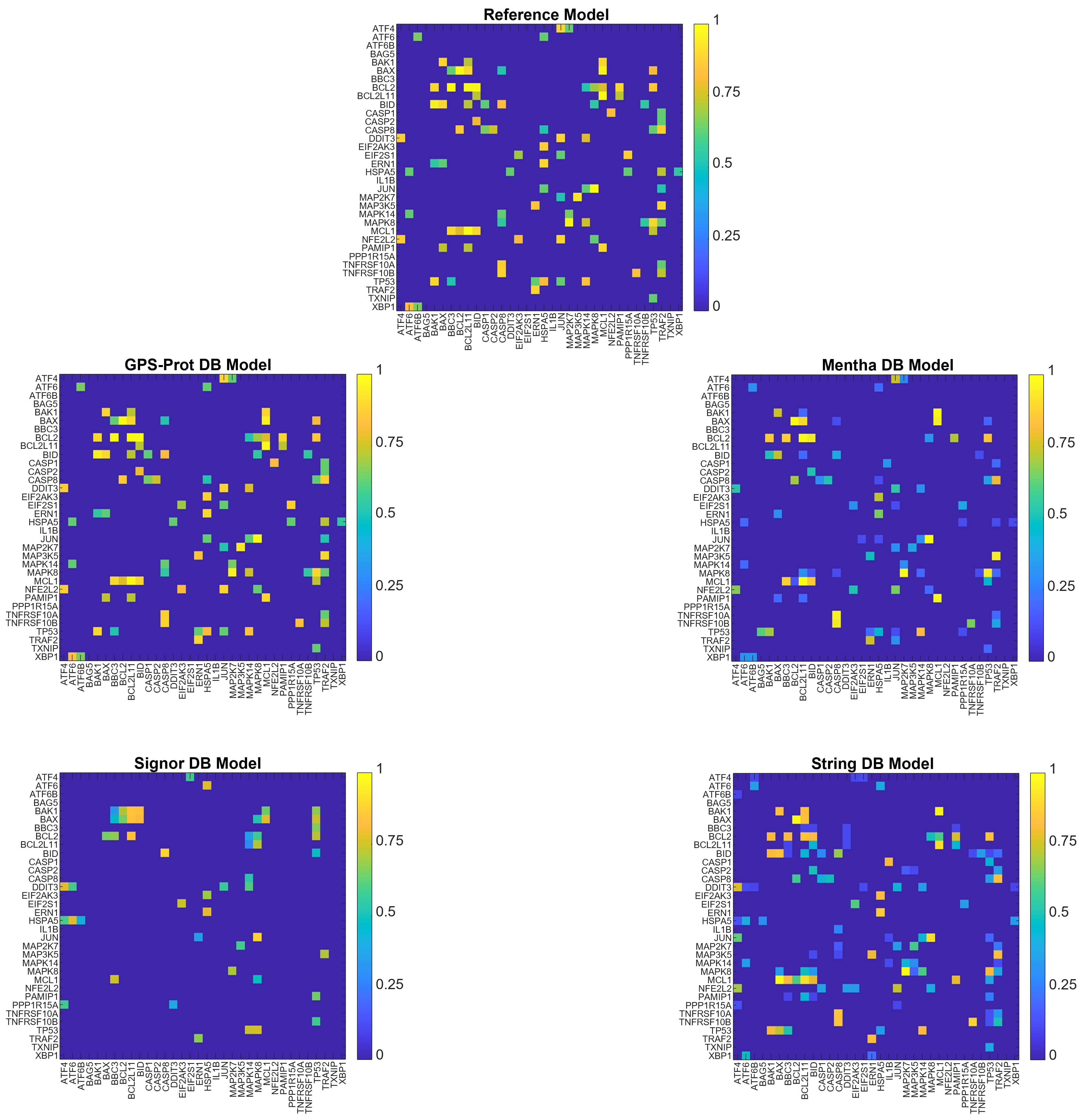}
    \caption{N34 UPR network models. The weights of interactions for each model are normalized with the maximum element of the matrix, and they are identified by the color bar. The (i, j) element of the matrix denotes the weight of the connection from j\textsuperscript{th} to i\textsuperscript{th} node.}
    \label{4}
\end{figure}

\paragraph{\textbf{N34 UPR+Apoptosis Network Models}}\label{UPR_Apopt_resdisc}
In Fig. \ref{4}, the graphical representations of N34 network models are shown. In Tabs. \ref{tab4.1} and \ref{tab4.2}, the nodes ranking scores and the number of MDN obtained for each database are reported, to be compared with results obtained for N34\textsubscript{mrf} model. \newline

As we can notice from Tab. \ref{tab4.2}, the dimension of MDN sets is very different across models, as already observed for N26 network models. Analogously, the lack of biological information affects the number of MDN for this network, in particular for N26\textsubscript{Sig} (see Tab. \ref{tab3.2}). \newline

The number of MDN found with GPS-Prot, Mentha, and String DBs is instead small enough to apply the algorithm for the identification of driver nodes, obtaining meaningful results (Tab. \ref{tab4.2}). \newline
\noindent These results are qualitatively quite different from what obtained for N10 and N15 networks, due to the new nodes and interactions. \newline
At variance with the other investigate networks, the MDN sets obtained from adjacency and weighted networks differ, for the same model. This is because of the heterogeneity of connections' weights respect to the total edges for a single node. In particular, with String DB we obtain two completely different MDN sets for the adjacency and weighted matrices, respectively, indicating the importance of the interaction scores in terms of biological information (see Tab. \ref{tab4.2}).\newline

In Tab. \ref{tab4.3}, the pairwise dGHD for N34 network models are reported. As already observed for the other networks, the smallest value of the distance is obtained with (GPS-Prot, Mentha) DB models. Due to the increasing size of the analyzed networks, the distance values also increase (see Tab. \ref{tab4.3}). Indeed, increasing the number of nodes can entail the inclusion of interactions not present in N34\textsubscript{rmf} model. The same consideration holds for every pair of models.

\newpage
\begin{table}[ht!]
    \centering
    \vspace*{-1 cm}
    Adjacency matrix (weighted matrix) \\
    $ $\\
    \resizebox{0.75\textwidth }{!}{\begin{tabular}{c | ccccc}
    \hline
    \hline
    Node & Reference & GPS-Prot & Mentha & Signor & String\\
    \hline
    ATF4&4 (-)      &  6 (8)  & 12 (13) & 2 (2)  &  4 (21)\\
    ATF6&4 (-)      &  6 (6)  & 12 (13) & 7 (22) &  4 (21)\\
    ATF6B&3 (-)     &  6 (8)  & 12 (13) & 7 (23) &  4 22\\
    BAG5&3 (-)      &  1 (1)  & 1 (1)   & 6 (20) &  5 (21)\\
    BAK1&3 (-)      &  6 (1)  & 1 (1)   & 6 (20) &  4 (16)\\
    BAX&3 (-)       &  6 (3)  & 3 (3)   & 6 (19) &  4 (16)\\
    BBC3&3 (-)      &  6 (5)  & 4 (5)   & 6 (21) &  4 (17)\\
    BCL2&3 (-)      &  6 (2)  & 2 (2)   & 6 (19) &  4 (16)\\
    BCL2L11&3 (-)   &  6 (4)  & 2 (4)   & 6 (19) &  4 (16)\\
    BID&3 (-)       &  6 (4)  & 3 (4)   & 6 (19) &  4 (16)\\
    CASP1&3 (-)     &  6 (6)  & 1 (1)   & 6 (21) &  4 (18)\\
    CASP2&3 (-)     &  6 (6)  & 1 (1)   & 6 (21) &  4 (18)\\
    CASP8&3 (-)     &  6 (6)  & 4 (5)   & 6 (19) &  4 (17)\\
    DDIT3&3 (-)     &  6 (7)  & 2 (2)   & 1 (1)  &  4 (18)\\
    EIF2AK3&7 (-)   &  7 (8)  & 14 (15) & 8 (22) &  4 22\\
    EIF2S1&5 (-)    &  1 (1)  & 13 (14) & 6 (20) &  4 (20)\\
    ERN1&3 (-)      &  6 (6)  & 10 (11) & 6 (20) &  5 (20)\\
    HSPA5&4 (-)     &  6 (6)  & 11 (12) & 6 (20) &  4 (20)\\
    IL1B&1 (-)      &  1 (1)  & 1 (1)   & 9 (22) &  4 (19)\\
    JUN&3 (-)       &  6 (6)  & 3 (3)   & 6 (19) &  4 (18)\\
    MAP2K7&4 (-)    &  6 (6)  & 7 (8)   & 6 (20) &  4 (19)\\
    MAP3K5&3 (-)    &  6 (8)  & 8 (9)   & 7 (21) &  4 (19)\\
    MAPK14&4 (-)    &  6 (6)  & 5 (7)   & 6 (20) &  4 (18)\\
    MAPK8&3 (-)     &  6 (6)  & 6 (7)   & 6 (20) &  4 (18)\\
    MCL1&3 (-)      &  6 (6)  & 3 (4)   & 6 (20) &  4 (16)\\
    NFE2L2&1 (-)    &  6 (8)  & 1 (1)   & 1 (1)  &  1 (1)\\
    PMAIP1&3 (-)    &  6 (6)  & 1 (1)   & 6 (20) &  4 (17)\\
    PPP1R15A&7 (-)  &  6 (7)  & 1 (1)   & 6 (21) &  5 (22)\\
    TNFRSF10A&3 (-) &  6 (7)  & 1 (1)   & 8 (20) &  4 (19)\\
    TNFRSF10B&3 (-) &  6 (6)  & 1 (1)   & 6 (19) &  4 (18)\\
    TP53&3 (-)      &  6 (5)  & 4 (6)   & 6 (19) &  4 (17)\\
    TRAF2&3 (-)     &  6 (6)  & 9 (10)  & 6 (19) &  4 (19)\\
    TXNIP&4 (-)     &  1 (1)  & 1 (1)   & 1 (1)  &  1 (1)\\
    XBP1&3 (-)      &  6 (7)  & 1 (1)   & 7 (21) &  4 (21)\\
    \hline
    \hline
    Eigenvalue & Reference & GPS-Prot & Mentha & Signor & String\\
    \hline
    0    &   6 (-)  &  6 (5)  &   7 (6)  &   17 (16)   &   3 (3)\\
    \hline
    \end{tabular}}
    \caption{Nodes ranking scores and eigenvalue for N34 UPR models.}
    \label{tab4.1}
\end{table}

\begin{table}[h!]
    \centering
    \addtolength{\leftskip} {-5cm}
    \addtolength{\rightskip}{-5cm}
    \resizebox{1.3\textwidth }{!}{
    \begin{tabular}{c | c | c}
    \hline
    \hline
    Model & MDN set (adj.) & MDN set (weight.)\\
    \hline
    N34\textsubscript{rmf}   & 6:  ATF6B, EIF2AK3, BBC3, MAP2K7, TNFRSF10, TXNIP    & -\\
    N34\textsubscript{GPS}   & 6: ATF6B, BAG5, BAK1, BBC3, IL1B, PPP1R15A           & 5: ATF6B, BAG5, BBC3, IL1B, PPP1R15A  \\
    N34\textsubscript{Men}   & 7: ATF6B, BAG5, BAK1, BBC3, BCL2L11, IL1B, PPP1R15A  & 6: ATF6B, ATF6, BAG5, BBC3, IL1B, PPP1R15A  \\
    N34\textsubscript{Str}   & 3: BAG5, ERN1, PPP1R15A                              & 3: ATF6B, BAG5, EIF2AK3   \\
    \hline
    \end{tabular}}
    \caption{MDN sets for N34 UPR models, referring to Tab. \ref{tab4.1}.}
    \label{tab4.2}
\end{table}

\clearpage
\begin{table}[ht]
    \centering
    dGHD (adjacency matrix)\\
    $ $\\
    \begin{tabular}{c | ccccc}
    \hline
    \hline
    Model & N34\textsubscript{mrf} & N34\textsubscript{GPS} & N34\textsubscript{Men} & N34\textsubscript{Sig} & N34\textsubscript{Str}\\
    \hline
    N34\textsubscript{mrf}    &     -    &   0.093 & 0.093  & \textbf{0.082} & 0.106\\
    N34\textsubscript{GPS}    &   0.093  &     -   & \textbf{0.020}  & 0.084 & 0.071\\
    N34\textsubscript{Men}    &   0.093  &   \textbf{0.020} &    -   & 0.081 & 0.070\\
    N34\textsubscript{Sig}    &   \textbf{0.082}  &   0.084 & 0.081  &   -   & 0.099\\
    N34\textsubscript{Str}    &   0.106  &   0.071 & 0.070  & 0.099 & -\\
    \hline
    \end{tabular}
    \caption{GHD distances between pairs of N34 UPR models. The most similar models are  N34\textsubscript{GPS}-N34\textsubscript{Men}; the database producing more similar results to the reference model is Signor.}
    \label{tab4.3}
\end{table}

\subsection{Statistical Analysis}

We analyse the rank distributions for the reference model (first columns of Tabs. \ref{tab1.1}, \ref{tab2.1} and \ref{tab4.1}), and the models obtained from each DB (other columns of Tabs. \ref{tab1.1}, \ref{tab2.1}, \ref{tab3.1} and \ref{tab4.1}) to highlights statistically significant differences and analogies.
We calculate the median and the percentile range for all the scores distributions of each built model. Results are reported  in Tab. \ref{tab1.1stat}. Obtained values show non-normal distributions of the ranking scores, a result that is statistically confirmed by the One-sample Kolmogorov-Smirnov test. \newline

The values for the median and percentile range (Tab. \ref{tab1.1stat}) calculated for String DB model are similar to those obtained for the reference model, and larger than values for the other three DB models. This is observed for the two smallest networks, N10 and N15. N26\textsubscript{Sig} model produces a distribution of ranking scores very different from the other N26 UPR models. The high number of obtained nodes with ranking score of 1 can be related to the fact the biological information regarding possible paired interactions involving these nodes is lacking in the database. For N34, models from Signor and String DBs seem to be very sensitive to the inclusion of weighted interactions; the values of median and percentile range drastically increase by considering the weighted matrix. This is an interesting result, highlighting the importance, at increasing network size, of a detailed biological information for a proper ranking score distribution description. Results remain almost unchanged when inserting the weighted information for the other two DB models (N34\textsubscript{GPS} and N34\textsubscript{Men}), probably because the weights are higher for these databases.

\clearpage\begin{table}[ht!]
    \centering
    Median and percentile range for adjacency matrix (weighted matrix) \\
    $ $\\
    \addtolength{\leftskip} {-5cm}
    \addtolength{\rightskip}{-5cm}
    \begin{tabular}[c]{ccccc}
    \hline
    \hline
    N10\textsubscript{mrf} & N10\textsubscript{GPS} & N10\textsubscript{Men} & N10\textsubscript{Sig} & N10\textsubscript{Str} \\
    \hline
    \textbf{8.0,  [7, 9]} & 3.5,  [1, 5] & 2.0,  [1, 4] & 3.0,  [1, 6] & \textbf{8.0,  [7, 8]}\\
    - & (3.5,  [1, 5]) & (2.0,  [1, 4]) & (3.0,  [1, 6]) & (\textbf{7.5,  [6, 8]}) \\
    \hline
    \hline
    N15\textsubscript{mrf} & N15\textsubscript{GPS} & N15\textsubscript{Men} & N15\textsubscript{Sig} & N15\textsubscript{Str} \\
    \hline
    \textbf{11.0, [10, 12]} & 9.0, [8, 10] & 6.0, [4, 9] & 3.0, [1.25, 6] & \textbf{11.0, [10, 12]}\\
    - & (9.0, [8, 10]) & (6.0, [4, 9]) & (3.0, [1.25, 6]) & (\textbf{11.0, [10, 12]}) \\
    \hline
    \hline
    - & N26\textsubscript{GPS} & N26\textsubscript{Men} & N26\textsubscript{Sig} & N26\textsubscript{Str} \\
    \hline
    - & 12.0, [10, 12] & 12.0, [1, 13] & 1.0,  [1, 6] & 14.0, [1, 15]\\
    - & (12.0, [10, 12]) & (12.0, [1, 13]) & (1.0,  [1, 6]) & (14.0, [1, 15]) \\
    \hline
    \hline
    N34\textsubscript{mrf} & N34\textsubscript{GPS} & N34\textsubscript{Men} & N34\textsubscript{Sig} & N34\textsubscript{Str} \\
    \hline
    \textbf{3.0, [3, 4]} & 6.0, [6, 6] & 3.0, [1, 8] & 6.0, [6, 6] & \textbf{4.0, [4, 4]}\\
    - & (6.0, [6, 7]) & (4.0, [1, 9]) & (30.0, [19, 21]) & (18.0, [17, 20]) \\
    \hline
    \end{tabular}
    \caption{Median and percentile range for all studied models.}
    \label{tab1.1stat}
\end{table}

 
We perform the Kruskal-Wallis test to search for at least one ranking scores' vector that is stochastically different from the others of the same network. \newline
In Tab. \ref{tab1.2stat}, we report results obtained from the Kruskal-Wallis test for paired DB models, and for the matrix containing all the distributions. The test gives: (i) no difference in terms of paired models, (ii) a significant difference when all the distributions are evaluated together. \newline

\begin{figure}[ht!]
   \centering
    \addtolength{\leftskip} {-5cm}
    \addtolength{\rightskip}{-5cm}
    \includegraphics[width=1.5\textwidth]{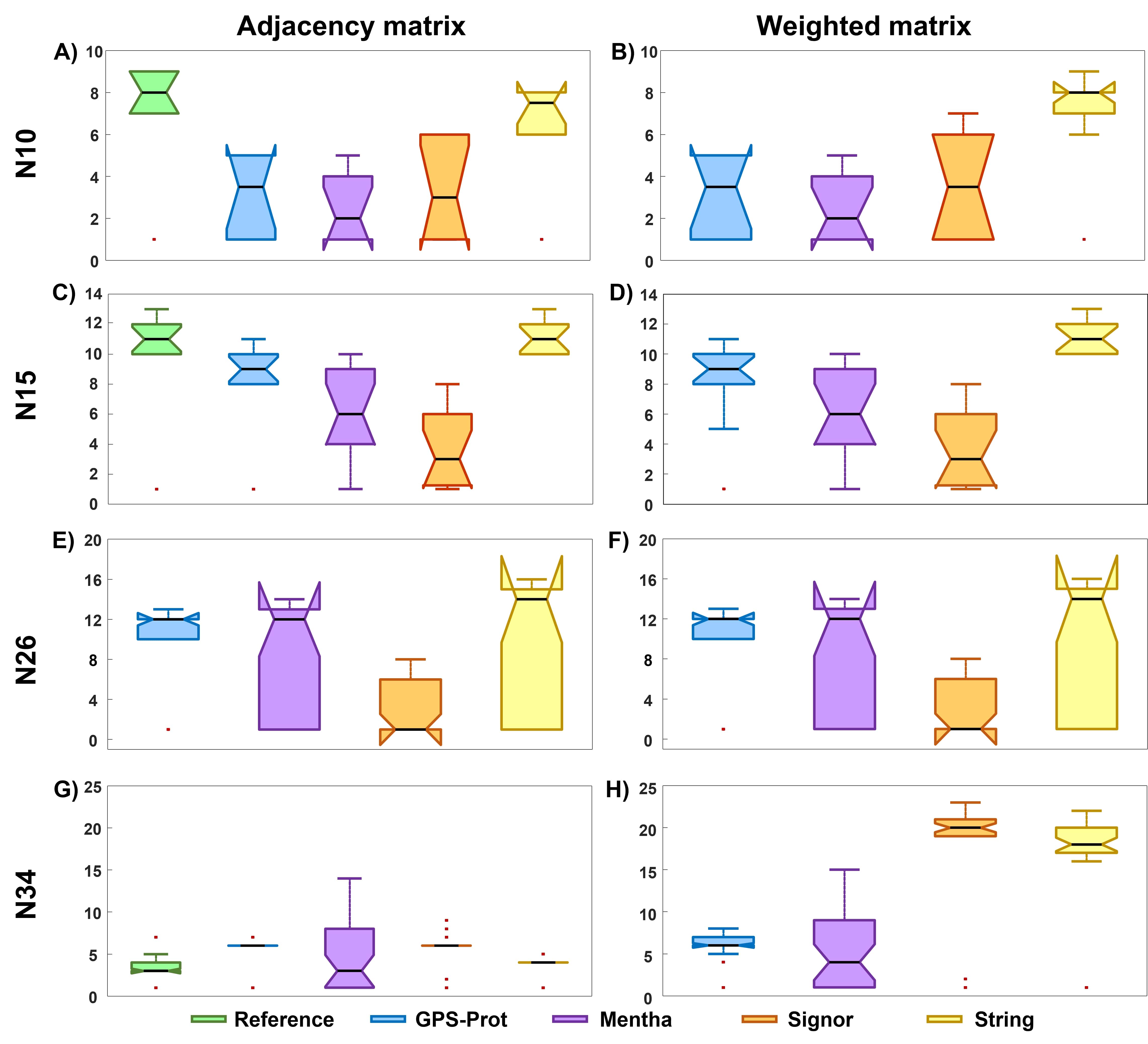}
     \caption{Kruskal-Wallis test of ranking score distributions of considered networks.}
    \label{ks_stat}
\end{figure}

In Fig. \ref{ks_stat} we plot the results of the Kruskal-Wallis test for all considered networks. Outlier points are individuated by red dots. In details, the outlier points individuated with the reference and N10\textsubscript{Str} models distributions (Fig. \ref{ks_stat} A-B) are related to the NFE2L2 node, which has 1 as the ranking score value for both adjacency and weighted models (upper panel of Tab. \ref{tab1.1}). The outlier points individuated by the test for the reference and N15\textsubscript{GPS} model (Fig. \ref{ks_stat} C-D) are related to NFE2L2, for the first model, and EIF2S1 and MAP3K5, for the latter one. The ranking scores change with respect to the N10 network models, since, for example, NFE2L2 is individuated as an outlier with a different DB model respect to what obtained with the previous networks; in this case, with String DB, the node possesses a score higher than 1, since NFE2L2 is outwardly connected to another node. The outlier points in Fig. \ref{ks_stat} E-F individuated by the test for N26\textsubscript{GPS} model are related to a small number of nodes (see Tab. \ref{tab3.1}), including once again NFE2L2 protein. For the case of N34 (Fig. \ref{ks_stat} G-H), there are many outlier points, associated (see Tab. \ref{tab4.1}) to nodes with ranking score values quite different from the median (see Tab. \ref{tab1.1stat}).

\clearpage\begin{table}[h!]
    \centering
    p-value for adjacency matrix (weighted matrix) \\
    $ $\\
    \begin{tabular}[c]{c | cccc}
    \hline
    \hline
    Model & N10\textsubscript{GPS} & N10\textsubscript{Men} & N10\textsubscript{Sig} & N10\textsubscript{Str}  \\
    \hline
    N10\textsubscript{GPS}    &   -  &   0.21 (0.21) & 0.52 (0.58)   & 0.46 (0.34)\\
    N10\textsubscript{Men}    &  0.21 (0.21)  &  -   & 0.51 (0.71)   & 0.25 (0.58)\\
    N10\textsubscript{Sig}    &  0.52 (0.58)  &   0.51 (0.71) & -  &  0.20 (0.80)\\
    N10\textsubscript{Str}    &  0.46 (0.34)  &   0.25 (0.58) & 0.20 (0.80)  & -\\
    \hline
    \hline
    $ $ & N15\textsubscript{GPS} & N15\textsubscript{Men} & N15\textsubscript{Sig} & N15\textsubscript{Str}\\
    \hline
    N15\textsubscript{GPS}    &   -                  &   0.06 (0.04) & 0.74 (0.97)   & 0.11 (0.048)\\
    N15\textsubscript{Men}    &  0.06 (0.04)         &  -                & 0.78 (1)   & 0.03 (0.03)\\
    N15\textsubscript{Sig}    &  0.74 (0.97)         &   0.78 (1.00) & -                   &  0.75 (0.40)\\
    N15\textsubscript{Str}    &  0.11 (0.048)         &   0.03 (0.03) & 0.75 (0.40)   & -\\
    \hline
    \hline
    $ $ & N26\textsubscript{GPS} & N26\textsubscript{Men} & N26\textsubscript{Sig} & N26\textsubscript{Str}\\
    \hline
    N26\textsubscript{GPS} & - & 0.02 (0.02) & 0.11 (0.18) & 0.36 (0.23)\\
    N26\textsubscript{Men} & 0.016 (0.02) & - & 0.16 (0.71) & 0.17 (0.10)\\
    N26\textsubscript{Sig} & 0.11 (0.18) & 0.16 (0.71) & - & 0.06 (0.04)\\
    N26\textsubscript{Str} & 0.36 (0.23) & 0.17 (0.10) & 0.06 (0.04) & -\\
    \hline
    \hline
    $ $ & N34\textsubscript{GPS} & N34\textsubscript{Men} & N34\textsubscript{Sig} & N34\textsubscript{Str}\\
    \hline
    N34\textsubscript{GPS}     & - & 0.21 (0.36) & 0.04 (0.11) & 0.12 (0.21)\\
    N34\textsubscript{Men}     & 0.22 (0.36) & - & 0.20 (0.15) & 0.18 (0.17)\\
    N34\textsubscript{Sig}     & 0.04 (0.11) & 0.20 (0.15) & - & 0.02 (0.045)\\
    N34\textsubscript{Str}     & 0.12 (0.21) & 0.18 (0.17) & 0.02 (0.045) & -\\
    \hline
    \end{tabular}
    \caption{Results of Kruskal-Wallis test for all studied models.}
    \label{tab1.2stat}
\end{table}

\begin{table}[h]
    \centering
    p-value for adjacency matrix (weighted matrix) \\
    $ $\\
    \begin{tabular}[c]{cccc}
    \hline
    \hline
    N10 & N15 & N26 & N34 \\
    \hline
    0.001 & $\ll$ 0.001 & $\ll$ 0.001 & $\ll$ 0.001 \\
    (0.001) & ($\ll$ 0.001) & ($\ll$ 0.001) & ($\ll$ 0.001)\\
    \hline
    \end{tabular}
    \caption{Results of Kruskal-Wallis test for all studied models.}
    \label{tab1.21stat}
\end{table}

In Tab. \ref{tab1.3stat}, we report results obtained from the signed-rank test for paired ranking score distributions of different combinations of models (same analysis done with the Generalized Hamming Distance), for all studied networks. The Wilcoxon signed rank test is also performed to compare the ranking score distributions obtained using databases with the distribution of the reference model, in the case of N10, N15, and N34 networks.

\clearpage

For the N10 network, all the models built on the four DBs are different from the reference model ($p <$ 0.05). Also, the String DB model is statistically different from the models built with the other three DBs. The test provides that there is no difference between the three GPS-Prot, Mentha, and Signor models, pair by pair ($p =$ 0.50 for N10\textsubscript{GPS}-N10\textsubscript{Men}, $p =$ 0.86 (0.59) for N10\textsubscript{GPS}-N10\textsubscript{Sig}, $p =$ 0.42 (0.30) for N10\textsubscript{Men}-N10\textsubscript{Sig}). \newline

For the N15 network, the signed rank test gives a bare difference for the distributions of Mentha and Signor DB ($p =$ 0.05 (0.06) for both pairs). The two-sided test provides that there is no difference between the reference model and N15\textsubscript{Str} model ($p =$ 1), as we can expect looking at ranking score distributions reported in Tab. \ref{tab2.1} \newline

For the N26 network, the two-sided test gives no difference in the distributions of GPS-Prot and Mentha DBs models ($p =$ 0.41). Since we do not built the reference model for the N26 network, it is not possible to perform the signed-rank test to compare it with the DB models. Much less we can perform the Wilcoxon rank sum test when switching from one network to another by changing the number of involved nodes. \newline

For the N34 network, the two-sided test gives no difference regarding GPS-Prot and Mentha DB models $(p =$ 0.41), as for N26, according to what is shown in Tab.  \ref{tab4.1}.

\begin{table}[h!]
    \centering
    \addtolength{\leftskip} {-5cm}
    \addtolength{\rightskip}{-5cm}
    p-value for adjacency matrix (weighted matrix)\\
    $ $\\
    \resizebox{1.4\textwidth }{!}{
    \begin{tabular}[c]{c | ccccc}
    \hline
    \hline
    Model & N10\textsubscript{mrf} & N10\textsubscript{GPS} & N10\textsubscript{Men} & N10\textsubscript{Sig} & N10\textsubscript{Str}\\
    \hline
    N10\textsubscript{mrf}     &  - & \textbf{0.004} (-) & \textbf{0.004} (-) & \textbf{0.004} (-) & \textbf{0.02} (-) \\
    N10\textsubscript{GPS}     &  \textbf{0.004} (-) & - & 0.50 (0.50) & 0.86 (0.59) & \textbf{0.004} (\textbf{0.004})\\
    N10\textsubscript{Men}     &  \textbf{0.004} (-) & 0.50 (0.50) & - & 0.42 (0.30, 0) & \textbf{0.004} (\textbf{0.004})\\
    N10\textsubscript{Sig}     &  \textbf{0.004} (-) & 0.86 (0.59) & 0.42 (0.30) & - & \textbf{0.004} (\textbf{0.004})\\
    N10\textsubscript{Str}     &  \textbf{0.016} (-) & \textbf{0.004} (\textbf{0.004}) & \textbf{0.004} (\textbf{0.004}) & \textbf{0.004} (\textbf{0.004}) & -\\
    \hline
    \hline
    $ $ & N15\textsubscript{mrf} & N15\textsubscript{GPS} & N15\textsubscript{Men} & N15\textsubscript{Sig} & N15\textsubscript{Str}\\
    \hline
    N15\textsubscript{mrf}     &  - & \textbf{0.005} (-) & \textbf{$\ll$ 0.001} (-) & \textbf{$\ll$ 0.001} (-) & 1.00 (-) \\
    N15\textsubscript{GPS}     &  \textbf{0.005} (-) & - & \textbf{$\ll$ 0.001} (\textbf{$\ll$ 0.001}) & \textbf{0.002} (\textbf{0.003}) & \textbf{$\ll$ 0.001} (\textbf{$\ll$ 0.001})\\
    N15\textsubscript{Men}     &  \textbf{$\ll$ 0.001} (-) & \textbf{$\ll$ 0.001} (\textbf{$\ll$ 0.001}) & - & 0.05 (0.06) & \textbf{$\ll$ 0.001} (\textbf{$\ll$ 0.001})\\
    N15\textsubscript{Sig}     &  \textbf{$\ll$ 0.001} (-) & \textbf{0.002} (\textbf{0.003}) & 0.05 (0.06) & - & \textbf{$\ll$ 0.001} (\textbf{$\ll$ 0.001})\\
    N15\textsubscript{Str}     &  1.00 (-) & \textbf{$\ll$ 0.001} (\textbf{$\ll$ 0.001}) & \textbf{$\ll$ 0.001}, 1 (\textbf{$\ll$ 0.001}) & \textbf{$\ll$ 0.001} (\textbf{$\ll$ 0.001}) & -\\
    \hline
    \hline
    Model & - & N26\textsubscript{GPS} & N26\textsubscript{Men} & N26\textsubscript{Sig} & N26\textsubscript{Str}\\
    \hline
    N26\textsubscript{GPS}& - & - & 0.41 (0.41) & \textbf{$\ll$ 0.001} (\textbf{0.003}) & \textbf{0.04} (\textbf{$\ll$ 0.001})\\
    N26\textsubscript{Men}& - & 0.41 (0.41) & - & \textbf{$\ll$ 0.001} (\textbf{$\ll$ 0.001}) & \textbf{0.03} (\textbf{0.02})\\
    N26\textsubscript{Sig}& - & \textbf{$\ll$ 0.001} (\textbf{0.003}) & \textbf{$\ll$ 0.001} (\textbf{$\ll$ 0.001}) & - & \textbf{$\ll$ 0.001} (\textbf{$\ll$ 0.001})\\
    N26\textsubscript{Str}& - & \textbf{0.041} (\textbf{$\ll$ 0.001}) & \textbf{0.03} (\textbf{0.02}) & \textbf{$\ll$ 0.001} (\textbf{$\ll$ 0.001}) & -\\
    \hline
    \hline
    $ $ & N34\textsubscript{mrf} & N34\textsubscript{GPS} & N34\textsubscript{Men} & N34\textsubscript{Sig} & N34\textsubscript{Str}\\
    \hline
    N34\textsubscript{mrf}& - & \textbf{$\ll$ 0.001} (-) & \textbf{$\ll$ 0.001} (-) & \textbf{$\ll$ 0.001} (-) & \textbf{0.006} (-)\\
    N34\textsubscript{GPS}& \textbf{$\ll$ 0.001} (-) & - & 0.41 (0.41) & \textbf{$\ll$ 0.001} (\textbf{0.003}) & \textbf{0.04} (\textbf{$\ll$ 0.001})\\
    N34\textsubscript{Men}& \textbf{$\ll$ 0.001} (-) & 0.41 (0.41) & - & \textbf{$\ll$ 0.001} (\textbf{$\ll$ 0.001}) & \textbf{0.03} (\textbf{0.02})\\
    N34\textsubscript{Sig}& \textbf{$\ll$ 0.001} (-) & \textbf{$\ll$ 0.001} (\textbf{0.003}) & \textbf{$\ll$ 0.001} (\textbf{$\ll$ 0.001}) & - & \textbf{$\ll$ 0.001} (\textbf{$\ll$ 0.001})\\
    N34\textsubscript{Str}& \textbf{0.006} (-) & \textbf{0.04} (\textbf{$\ll$ 0.001}) & \textbf{0.03} (\textbf{0.02}) & \textbf{$\ll$ 0.001} (\textbf{$\ll$ 0.001}) & -\\
    \hline
    \end{tabular}}
    \caption{Results of Wilcoxon signed rank test for all studied models.}
    \label{tab1.3stat}
\end{table}

Finally, we perform the Wilcoxon rank sum test, by comparing the score distributions of N10, N15 and N34 UPR models, also including the reference models, to evaluate the importance of adding new nodes to the networks. We do not include N26 in this analysis because we do not have the reference model for N26. Results are reported in Tab. \ref{tab1.4stat}.

\begin{table}[h!]
    \centering
    p-value for adjacency matrix (weighted matrix)\\
    $ $\\
    \begin{tabular}[c]{ccccc}
    \hline
    \hline
     N15\textsubscript{mrf} & N15\textsubscript{GPS} & N15\textsubscript{Men} & N15\textsubscript{Sig} & N15\textsubscript{Str}\\
    \hline
    \textbf{$\ll$ 0.001} & \textbf{0.001} & \textbf{0.01} & 0.65 & \textbf{$\ll$ 0.001}\\
    (-) & (\textbf{0.002}) &(\textbf{0.01}) &(0.84) &(\textbf{$\ll$ 0.001}) \\
    \hline
    \end{tabular}
    $ $\\
    $ $\\
    Comparison with N10 UPR models
    $ $\\
    \begin{tabular}[c]{ccccc}
    \hline
    \hline
    N34\textsubscript{mrf} & N34\textsubscript{GPS} & N34\textsubscript{Men} & N34\textsubscript{Sig} & N34\textsubscript{Str}\\
    \hline
    \textbf{$\ll$ 0.001} &\textbf{$\ll$ 0.001} &0.30 &\textbf{$\ll$ 0.001} &\textbf{$\ll$ 0.001}\\
    (-) &(\textbf{$\ll$ 0.001}) &(0.50) &(\textbf{$\ll$ 0.001}) &(\textbf{$\ll$ 0.001})\\
    \hline
    \end{tabular}
    $ $\\
    Comparison with N15 UPR models
    $ $\\
    \begin{tabular}[c]{ccccc}
    \hline
    \hline
    N34\textsubscript{mrf} & N34\textsubscript{GPS} & N34\textsubscript{Men} & N34\textsubscript{Sig} & N34\textsubscript{Str}\\
    \hline
    \textbf{$\ll$ 0.001} &\textbf{$\ll$ 0.001} &0.86 &\textbf{0.005} &\textbf{$\ll$ 0.001}\\
    (-) &(\textbf{$\ll$ 0.001}) &(0.57) &(\textbf{$\ll$ 0.001}) &(\textbf{$\ll$ 0.001})\\
    \hline
    \end{tabular}
    \caption{Results of Wilcoxon rank sum test for N15 UPR models with N10, and for N34 UPR models with N10 and N15.}
    \label{tab1.4stat}
\end{table}

As we can note from Tab. \ref{tab1.4stat}., the test provides that there is no difference ($p = 0.65$ (0.84)) between the distributions regarding Signor DB model, passing from the N10 to the N15 network. So we can assert that the model produced by Signor DB is not affected by changes in the dimensionality of the network. In particular, the test gives a significant difference regarding String DB model ($p\ll0.001$). For N34, the test shows that there is no difference between the distributions of ranking scores regarding Mentha DB model, by comparing N34\textsubscript{Men} model with N10\textsubscript{Men} ($p = 0.30$ (0.50)) and N15\textsubscript{Men} models ($p = 0.86$ (0.57)). We can assert that, in this case, the N34 model produced by Mentha DB is not affected by changes in the dimensionality of the network. Also for N34 network, the test gives a significant difference regarding GPS-Prot and String DB models ($p\ll0.001$).

\section{Discussion on the Biological Significance of Structural Controllability Results}\label{bio_disc}
\noindent Referring to ranking scores distributions and MDN sets results reported  in Tabs. \ref{tab1.1} and \ref{tab1.2}, respectively, we find that, for all N10 DB models, ATF6 and EIF2AK3 are identified as driver nodes. This is an interesting finding because the ER stress sensors can be considered among the most important proteins involved in the UPR mechanism, since the adaptive response starts from them. Furthermore, in the case of the adjacency N10\textsubscript{Str} model, the driver nodes coincide exactly with the three ER sensors. Despite the interesting results for the N10 models, the dimension of the network is probably still too small to obtain very significant biological information about the mechanism.\newline

Going to the N15 network models, in most cases, two of the ER stress sensors, i.e., ATF6 and ERN1, are identified as driver nodes, confirming the results obtained with the N10 network (see Tabs. \ref{tab1.2} and \ref{tab2.2}). The same considerations done for the N10 network hold also for the N15 network, with the control theory results emphasizing the importance of the ER stress sensors in the UPR mechanism (Tab. \ref{tab2.2}). In particular, the ERN1-TRAF2-ASK1 complex is included in the N15 network, which is of fundamental importance for the UPR and apoptosis mechanisms, especially in neuronal cells \cite{ZENG2015530, nishitoh2002ask1}. Other nodes included in the N15 MDN sets, for all models, are MAPK proteins. In particular, MAPK14 is a very important protein because it also constitutes the main p38 MAPK activity in the heart, and it is responsible for the promotion of myocyte apoptosis via downstream targets DDIT3 and TP53 \cite{Sharov2003, Marber2010,Eiras2006,Fiordaliso2001}.\newline

Regarding the N34, the MDN sets results (Tab. \ref{tab4.2}) obtained for the DB models are very different from the set of the N34\textsubscript{mrf} model, 
except for activating transcription factor 6 $\beta$ ATF6B, which results as DN in all sets. It has a survival role in the ER stress response in pancreatic $\beta$-cells, and ATF/CREB is involved in diabetes \cite{THUERAUF200722865,ODISHO2015111,BILEKOVA202185,Tchapyjnikov2010}, as a master regulator for hepatic gluconeogenesis \cite{Cui2021}.
It is not straightforward to attribute a biological meaning to the set of driver nodes we obtain, and to directly relate them to DNs for smaller networks. The inclusion of a complete new mechanism (the apoptosis) in N34, and thus the increase in network dimension with respect to N15 (number of nodes more than doubled, number of edges almost increased fivefold) may be responsible for the discrepancies in obtained DN sets. By increasing the number of nodes and edges, there could be the possibility of losing biological information in the models because the amount of missing information stored in the databases increases.
Nevertheless, PPP1R15A is a driver node for N10, N15 and N34. It is a gene involved in facilitating recovery of cells from stress \cite{Choy2015,Santos2016}.\newline

We are not able to extract biological information from structural controllability results for N26 network models because they are not very significant; in fact, about the 50\% or more nodes result as drivers. For this reason, we do not carried out a more thorough analysis as we do for the other examined networks.

\section{Conclusions} \label{concl}

\noindent The Unfolded Protein Response is one of the most important biological mechanism in living systems which take place in the endoplasmic reticulum, whose functionality is to restore the physiological homeostatic state of cells \cite{Hetz2012, SHEN2002, SHEN2005, Almanza, Walter2011, HETZ2018,GESSNER2014}. A failure in the efficiency of the ER stress adaptive response causes a more or less severe accumulation of unfolded or misfolded proteins, which are known to create aggregates that can damage cells and tissues. This condition is the main reason for the development of amyloidosis and neurodegenerative diseases \cite{Scheper2015, Vidal2012, ijms21176127, biom10081090}. \newline

The structural controllability analysis, based on Kalman's rank condition \cite{Kalman1963}, provides a very useful tool to understand the behaviour of complex systems, which can be described as networks characterized by nodes and edges. The Minimum Drive Nodes approach to studying a complex network can provide important biological information on the proteins involved in a certain mechanism. By applying this theory to the investigated network models, it is possible to identify which are the fundamental nodes that control all the network \cite{Liu2011_1, 8028698}. \newline

Nevertheless, our study, based on experimentally built networks, can be used for a reliable prediction of the development of dysfunction in biological processes by varying nodes and edges in the related networks, for which it is necessary to know a significant piece of biological information. For this purpose, we try to apply the structural controllability analysis to the UPR mechanism by studying four different networks, based on the biological information from different protein-protein interaction databases. We point out that for this method to be effective, sufficient information must be stored in PPI experimental databases. \newline

Our results nicely confirm the hypothesis that the ER stress sensors are among the most important participants in the UPR adaptive response, even when a part of the biological information is missing in the network model. This can mean that a dysfunction in the ER stress sensors pathways, in particular in the induction of chaperones and ER-associated degradation, could be disastrous for the living systems \cite{Junjappa2018, Lin2008, Amen2019, Bhattarai2020}. Further, our study also has the potential to uncover novel regulators of UPR and other key mechanisms, potentially associated with multiple dysfunctions and diseases \cite{ZENG2015530, nishitoh2002ask1, Sharov2003, Marber2010,Eiras2006,Fiordaliso2001, THUERAUF200722865,ODISHO2015111,BILEKOVA202185,Tchapyjnikov2010, Cui2021, Choy2015,Santos2016}.

\section*{Author contributions}
\noindent Conceptualization, N.L., A.L., L.C., S.F.; methodology, N.L., A.L., M.A.G.M., L.C.; investigation, N.L.; formal analysis, N.L.; writing--original draft preparation, N.L. and L.C.; writing--review and editing, all authors; visualization, N.L., M.A.G.M, L.C.; supervision, A.L., L.C, S.F . All authors have read and agreed to the published version of the manuscript.

\section*{Acknowledgments}
\noindent A.L., L.C. and S.F. acknowledge the support of the International Center for Relativistic Astrophysics Network (ICRANet). All the authors acknowledge the support of the Italian National Group for Mathematical Physics (GNFM-INdAM). All the authors acknowledge the support from HORIZON-CL4-2021-DIGITAL-EMERGING-01 –``Muquabis" project and the EBRAINS research infrastructure. We thank Dr. Alberto Luini for useful scientific discussions.

\section*{Abbreviations}
\noindent The following abbreviations are used in the manuscript:\\

\noindent 
\begin{tabular}{@{}ll}
UPR & unfolded protein response\\
ER & endoplasmic reticulum\\
PPI & protein-protein interaction/s\\
MDN & minimum driver nodes\\
GHD & Generalized Hamming Distance\\
DB / DBs & database / databases\\
\end{tabular}

\bibliographystyle{elsarticle-num}

\end{document}